\documentclass[12pt, preprint]{aastex}
\usepackage{epstopdf}
\usepackage{graphicx,epsfig}
\usepackage{natbib}
\usepackage{float}
\usepackage[toc,page]{appendix}
\usepackage{longtable}
\usepackage[normalem]{ulem}

\newcommand{\numberCatalogEBs}{2878}

\newcommand{\rev}{{\bf }}

\begin{document}

\title{\emph{Kepler} Eclipsing Binary Stars. VII. The Catalog of Eclipsing Binaries Found in the Entire Kepler Data-Set.}

%Revised \today

\author{Brian Kirk}
\affil{National Radio Astronomy Observatory, North American ALMA Science Center, 520 Edgemont Road, Charlottesville, VA 22903}
\affil{Villanova University, Dept. of Astrophysics and Planetary Science, 800 E Lancaster Ave, Villanova, PA 19085}
\email{bkirk@nrao.edu}

\author{Kyle Conroy}
\affil{Vanderbilt University, Dept. of Physics and Astronomy, VU Station B 1807, Nashville, TN 37235}
\affil{Villanova University, Dept. of Astrophysics and Planetary Science, 800 E Lancaster Ave, Villanova, PA 19085}
\email{kyle.conroy@vanderbilt.edu}

\author{Andrej Pr\v sa}
\affil{Villanova University, Dept. of Astrophysics and Planetary Science, 800 E Lancaster Ave, Villanova, PA 19085}
\email{aprsa@villanova.edu}

\author{Michael Abdul-Masih}
\affil{Rensselaer Polytechnic Institute, Dept. of Physics, Applied Physics, and Astronomy, 110 8th St, Troy, NY 12180}
\affil{Villanova University, Dept. of Astrophysics and Planetary Science, 800 E Lancaster Ave, Villanova, PA 19085}

\author{Angela Kochoska}
\affil{Faculty of Mathematics and Physics, University of Ljubljana, Jadranska 19, 1000 Ljubljana, Slovenia}

\author{Gal Matijevi\v c}
\affil{Villanova University, Dept. of Astrophysics and Planetary Science, 800 E Lancaster Ave, Villanova, PA 19085}

\author{Kelly Hambleton}
\affil{Jeremiah Horrocks Institute, University of Central Lancashire, Preston, PR1~2HE}

%_____________Tier 2______________________%

\author{Thomas Barclay}
\affil{NASA Ames Research Center/BAER Institute, Moffett Field, CA 94035, USA}

\author{Steven Bloemen}
\affil{Department of Astrophysics/IMAPP, Radboud University Nijmegen, 6500 GL Nijmegen, The Netherlands}

\author{Tabetha Boyajian}
\affil{J.W. Gibbs Laboratory, Yale University, 260 Whitney Avenue, New Haven, CT 06511}

\author{Laurance R. Doyle}
\affil{IMoP at Principia College, Elsah, Illinois  62028}
\affil{SETI Institute, 189 Bernardo Ave. Mountain View, California 94043} 

\author{B.J. Fulton}
\affil{Las Cumbres Observatory Global Telescope Network, Goleta, CA 93117, USA}

\author{Abe Johannes Hoekstra}
\affil{planethunters.org}

\author{Kian Jek}
\affil{planethunters.org}

\author{Stephen R. Kane}
\affil{San Francisco State University, 1600 Holloway Avenue, San Francisco, CA 94132}

\author{Veselin Kostov}
\affil{Department of Astronomy \& Astrophysics at U of T, Toronto, Ontario, Canada M5S 3H4}

\author{David Latham}
\affil{Harvard-Smithsonian Center for Astrophysics, 60 Garden Street, Cambridge, MA 02138}

\author{Tsevi Mazeh}
\affil{Wise Observatory, Tel Aviv University, Tel Aviv, Israel}

\author{Jerome A.\ Orosz} 
\affil{San Diego State University, 5500 Campanile Dr., San Diego, CA 92182}

\author{Joshua Pepper}
\affil{Lehigh University, Department of Physics, 16 Memorial Drive East, Bethlehem, PA 18015}

\author{Billy Quarles}
\affil{\affil{NASA Ames Research Center, Astrobiology and Space Science Division MS 245-3, Moffett Field, CA 94035}}

\author{Darin Ragozzine}
\affil{Florida Institute of Technology, Physics and Space Sciences, 150 W. University Blvd. Melbourne, FL 32901}

\author{Avi Shporer}
\affil{Jet Propulsion Laboratory, California Institute of Technology, 4800 Oak Grove Drive, Pasadena, CA 91109, USA}
\affil{Sagan Fellow}

\author{John Southworth}
\affil{Astrophysics Group, Keele University, Staffordshire, ST5 5BG, UK}

\author{Keivan Stassun}
\affil{Vanderbilt University, Nashville, Tennessee 37240}

\author{Susan E.\ Thompson}
\affil{SETI Institute/NASA Ames Research Center, Moffett Field, CA 94035}

\author{William F.\ Welsh}
\affil{San Diego State University, 5500 Campanile Dr., San Diego, CA 92182}

%________________Tier 3______________________%

\author{Eric Agol}
\affil{Astronomy Department, University of Washington, Seattle, WA 98195}

\author{Aliz Derekas}
\affil{ELTE Gothard Astrophysical Observatory, H-9704 Szombathely, Szent Imre herceg u. 112, Hungary}
\affil{Konkoly Observatory, Research Centre for Astronomy and Earth Sciences, Hungarian Academy of Sciences, H-1121}

\author{Jonathan Devor}
\affil{Tel Aviv University, Department of Astrophysics, Tel Aviv 69978, Israel}

\author{Debra Fischer}
\affil{Yale University, New Haven, CT 06520-8101 USA}

\author{Gregory Green}
\affil{Harvard-Smithsonian Center for Astrophysics, 60 Garden Street, MS-10, Cambridge, MA 02138}

\author{Jeff Gropp}
\affil{Villanova University, Dept. of Astrophysics and Planetary Science, 800 E Lancaster Ave, Villanova, PA 19085}

\author{Tom Jacobs}
\affil{planethunters.org}

\author{Cole Johnston}
\affil{Villanova University, Dept. of Astrophysics and Planetary Science, 800 E Lancaster Ave, Villanova, PA 19085}

\author{Daryll Matthew LaCourse}
\affil{planethunters.org}

\author{Kristian Saetre}
\affil{planethunters.org}

\author{Hans Schwengeler}
\affil{Astronomisches Institut der Universitat Basel, Venusstrasse 7, CH-4102 Binningen, Switzerland}

\author{Jacek Toczyski}
\affil{University of Virginia, 4040 Lewis and Clark Dr, Charlottesville, VA 22911}

\author{Griffin Werner}
\affil{Villanova University, Dept. of Astrophysics and Planetary Science, 800 E Lancaster Ave, Villanova, PA 19085}

\author{Matthew Garrett, Joanna Gore, Arturo O. Martinez, Isaac Spitzer, Justin Stevick, Pantelis C. Thomadis, Eliot Halley Vrijmoet, and Mitchell Yenawine}
\affil{San Diego State University, 5500 Campanile Dr., San Diego, CA 92182}

\author{Natalie Batalha}
\affil{San Jose State University, One Washington Square, San Jose, CA 95192}

\author{William Borucki}
\affil{NASA Ames Research Center, Moffett Field, CA 94035, USA}

\begin{abstract}
The primary \emph{Kepler} Mission provided nearly continuous monitoring of $\sim$200,000 objects with unprecedented photometric precision. We present the final catalog of eclipsing binary systems within the 105 deg$^2$ \emph{Kepler} field of view. This release incorporates the full extent of the data from the primary mission (Q0-Q17 Data Release). As a result, new systems have been added, additional false positives have been removed, ephemerides and principal parameters have been recomputed, classifications have been revised to rely on analytical models, and eclipse timing variations have been computed for each system. We identify several classes of systems including those that exhibit tertiary eclipse events, systems that show clear evidence of additional bodies, heartbeat systems, systems with changing eclipse depths, and systems exhibiting only one eclipse event over the duration of the mission. We have updated the period and galactic latitude distribution diagrams and included a catalog completeness evaluation. The total number of identified eclipsing and ellipsoidal binary systems in the \emph{Kepler} field of view has increased to \numberCatalogEBs{}, 1.3\% of all observed \emph{Kepler} targets. An online version of this catalog with downloadable content and visualization tools is maintained at \texttt{http://keplerEBs.villanova.edu}.
\end{abstract}

\section{Introduction}
The contribution of binary stars and, in particular, eclipsing binaries (EBs) to astrophysics cannot be overstated. EBs can provide fundamental mass and radius measurements for the component stars (e.g., see the extensive review by Andersen 1991). These mass and radius measurements in turn allow for accurate tests of stellar evolution models (e.g., \citealt{pols:1997}; \citealt{schroder:1997}; \citealt{guinan:2000}; \citealt{torres:2002}). In cases where high-quality radial velocity measurements exist for both stars in an EB, the luminosities computed from the absolute radii and effective temperatures can lead to a distance determination. Indeed, EBs are becoming widely used to determine distances to the Magellanic Clouds, M31, and M33 (\citealt{guinan:1998}; \citealt{ribas:2002}; \citealt{wyithe:2001, wyithe:2002}; \citealt{hilditch:2005}, \citealt{bonanos:2003, bonanos:2006}; \citealt{north:2010}).

%Paczynski \& Sasselov 1997; Paczynski \& Pojmanski 2000;

Large samples are useful to determine statistical properties and for finding rare binaries which may hold physical significance (for example, binaries with very low mass stars, binaries with stars in short-lived stages of evolution, very eccentric binaries that show large apsidal motion, etc). Catalogs of EBs from ground-based surveys suffer from various observational biases such as limited accuracy per individual measurement, complex “window” functions (e.g., observations from ground based surveys can only be done during nights with clear skies and during certain seasons). 

The NASA \emph{Kepler} Mission, launched in March 2009, provided essentially uninterrupted, ultra-high precision photometric coverage of $\sim$200,000 objects within a 105 deg$^2$ field of view in the constellations of Cygnus and Lyra for four consecutive years. The details and characteristics of the \emph{Kepler} instrument and observing program can be found in  \citet{batalha:2010, borucki:2011, caldwell:2010, koch:2010}. The mission has revolutionized the exoplanetary and eclipsing binary field. The previous catalogs can be found in \citet[hereafter Paper I]{prsa:2011} and \citet[hereafter Paper II]{slawson:2011} at \texttt{http://keplerEBs.villanova.edu/v1} and \texttt{http://keplerEBs.villanova.edu/v2}, respectively. The current catalog and information about its functionality is available at \texttt{http://keplerEBs.villanova.edu}. 

\section{The Catalog}
The \emph{Kepler} Eclipsing Binary Catalog lists the stellar parameters from the \emph{Kepler} Input Catalog (KIC) augmented by: primary and secondary eclipse depth, eclipse width, separation of eclipse, ephemeris, morphological classification parameter, and principal parameters determined by geometric analysis of the phased light curve. 

The online Catalog also provides the raw and detrended data for $\sim$30min (long) cadence, and raw $\sim$1min (short) cadence data (when available), an analytic approximation via a polynomial chain (polyfit; \citealt{prsa:2008}), and eclipse timing variations (ETV) (\citet{conroy:2014}, hereafter Paper IV and Orosz 2015 et al., in preparation). The construction of the Catalog consists of the following steps (1) EB signature detection (Section \ref{sec:CatAdd}); (2) data detrending: all intrinsic variability (such as chromospheric activity, etc) and extrinsic variability (i.e. third light contamination and instrumental artifacts) are removed by the iterative fitting of the photometric baseline \citep{prsa:2011}; (3) the determination of the ephemeris: the time-space data are phase-folded and the dispersion minimized; (4) Determination of ETVs (Section \ref{sec:ETV}); (5) analytic approximation: every light curve is fit by a polyfit \citep{prsa:2008}; (6) morphological classification via Locally Linear Embedding (LLE; Section \ref{sec:LLE}), a nonlinear dimensionality reduction tool is used to estimate the ``detachedness'' of the system (\citealt{matijevic:2012}, hereafter Paper III); (7) EB characterization through geometric analysis and (8) diagnostic plot generation for false positive (FP) determination. Additional details on these steps can be found in Paper I, Paper II, Paper III, and Paper IV. For inclusion in this Catalog we accept bonafide EBs and systems that clearly exhibit binarity through photometric analysis (heartbeats and ellipsoidals (Section \ref{sec:HB}). Throughout the Catalog and online database we use a system of subjective flagging to label and identify characteristics of a given system that would otherwise be difficult to validate quantitatively or statistically. Examples of these flags and their uses can be seen in Section 8. Although best efforts have been taken to provide accurate results, we caution that not all systems marked in the Catalog are guaranteed to be EB systems. There remains the possibility that some grazing EB signals may belong to small planet candidates or are contaminated by non-target EB signals. An in-depth discussion on Catalog completeness is presented in Section 10.

In this release, we have updated the Catalog in the following ways: 
\begin{enumerate}
\item The light curves of \emph{Kepler} Objects of Interest (KOIs) once withheld as possibly containing planetary transit events but since rejected have been included.
 
\item An increased baseline allowed for ETVs to be determined and therefore a greater precision of all ephemerides \citep{orosz:2012, conroy:2014}. Systems previously having indeterminable periods were re-examined and included, if additional eclipses were observed. 

\item Period \rev{and $BJD_0$ error estimates} are provided across the Catalog. The period error analysis is derived from error propagation theory and applied through an adaptation of the Period Error Calculator algorithm of \citet{mighell:2013} and $BJD_0$ errors are estimated by fitting a gaussian to the bisected primary eclipse.

\item Quarter Amplitude Mismatch (QAM) systems have been rectified by scaling the affected season(s) to the season with the largest amplitude; a season being one of the four rotations per year to align the solar arrays. Once corrected, the system was reprocessed by the standard pipeline. 

\item An additional 13 systems identified by the independent Eclipsing Binary Factory pipeline (Section \ref{sec:EBF}) have been processed and added to the Catalog.

%\item An additional X systems identified by the independent Robovetter (Section Y) have been processed and added to the Catalog.

\item Additional systems identified by Planet Hunters (Section \ref{sec:PH}), a citizen science project that makes use of the Zooniverse toolset to serve flux-corrected light curves from the public \emph{Kepler} data-set, have been added. 

\item All systems were investigated to see if the signal was coming from the target source or a nearby contaminating signal. If a target was contaminated, the contaminated target was removed and the real source, if a \emph{Kepler} target, was added to the Catalog (Section \ref{sec:falseEB}). A follow-up paper (Abdul-Masih et al., in preparation) will address the EBs that are not Kepler targets, but whose light curves can be recovered from target pixel files to a high precision and fidelity.

\item Long-cadence exposure causes smoothing of the light curves due to $\sim$30-min integration times (Section \ref{sec:LCD}). Deconvolution of the phased long-cadence data polyfit allowed removal of integration smoothing, resulting in a better representation of the actual EB signal.

\item Classifications were done via Locally Linear Embedding (LLE, Section \ref{sec:LLE}), a general nonlinear dimensionality reduction tool, to give a value between 0 and 1, to represent detachedness as in \citet{matijevic:2012}. 

\item Visualization of the data-set through tSNE (Maaten \& Hinton 2008) has been provided to reveal global and local structure of similarity between systems (Section \ref{sec:tSNE}).

\item Threshold Crossing Events have been manually vetted for additional EB signals resulting in additions to the Catalog.

\item Follow-up data are provided where applicable. This may consist of spectroscopic data (Section \ref{sec:spectra}) or additional follow-up photometric data.

\end{enumerate}

\section{Catalog Additions}
\label{sec:CatAdd}
The previous release of the Catalog (Paper II) contained 2165 objects with eclipsing binary and/or ellipsoidal variable signatures, through the second \emph{Kepler} data release (Q0-Q2). In this release, \numberCatalogEBs{} objects are identified and analyzed from the entire data-set of the primary \emph{Kepler} mission (Q0-Q17). All transit events were identified by the main \emph{Kepler} pipeline \citep{jenkins:2002, jenkins:2010c} and through the following sources explained here. 

%Any systems added to the Catalog were cross-checked against the list of chromospherically active stars \citep{basri:2011, coughlin:2010}, low-mass binaries, short-period pulsators \citep{debosscher:2011}, and a GO-reported list \citep{morrison:2011}. Other cross-checks have been performed, namely against instrumental systematics, safe mode events, and guide star artifacts. 

%Removing this stemmed from comments on what this was for, surely there are chromospherically active EBs, pulsating EBs, etc. So was this meant for TCE vetting? Removing this section also clears up doing these cross-checks.

\subsection{Rejected KOI Planet Candidates}
\label{sec:KOI_FP}
The catalog of KOIs \citep{mullally:2015} provides a list of detected planets and planet candidates. There is an inevitable overlap in attributing transit events to planets, severely diluted binaries, low mass stellar companions, or grazing eclipsing binaries. As part of the \emph{Kepler} working group efforts, these targets are vetted for any EB-like signature, such as depth changes between successive eclipses (the so-called even-odd culling), detection of a secondary eclipse that is deeper than what would be expected for an R $\textless$ 2R$_{Jup}$ planet transit (occultation culling), hot white dwarf transits (\citet{rowe:2010}; white dwarf culling), spectroscopic follow-up where large amplitudes or double-lined spectra are detected (follow-up culling), and automated vetting programs known as ``robovetters'' (Coughlin et al. 2015, in prep.). High-resolution direct imaging (AO and speckle) and photo-center centroid shifts also indicate the presence of background EBs. These systems can be found in the online \emph{Kepler} Eclipsing Binary catalog by searching the ``KOI'' flag.

\subsection{Eclipsing Binary Factory}
\label{sec:EBF}
The Eclipsing Binary Factory (EBF) \citep{parvizi:2014} is a fully automated, adaptive, end-to-end  computational pipeline used to classify EB light curves. It validates EBs through an independent neural network classification process. The EBF uses a modular approach to process large volumes of data into patterns for recognition by the artificial neural network. This is designed to allow archival data from time-series photometric surveys to be direct input, where each module's parameters are tunable to the characteristics of the input data (e.g.,  photometric precision, data collection cadence, flux measurement uncertainty) and define the output options to produce the probability that each individual system is an EB. A complete description can be found in \cite{stassun:2013}. The neural network described here was trained on previous releases of this Catalog. The EBF identified 68 systems from Quarter 3 data. Out of the 68 systems submitted to our pipeline only 13 were validated and added to the Catalog.

\subsection{Planet Hunters}
\label{sec:PH}
Planet Hunters is a citizen science project \citep{fischer:2012} that makes use of the Zooniverse toolset (Lintott et al. 2008 and Smith et al. 2010) to serve flux-corrected light curves from the \emph{Kepler} public release data. This process is done manually by visual inspection of each light curve for transit events. For a complete description of the process see \citet{fischer:2012}. Identified transit events not planetary in nature are submitted to our pipeline for further vetting and addition to the Catalog. 

\subsection{Increased Baseline Revisions} 
An increased timespan allows the ephemerides for all EB candidates to be determined to a greater precision. All ephemerides have been manually vetted, but for certain systems it was impossible to uniquely determine the periods, i.e., for systems with equal depth eclipses (versus a single eclipse at half-period). To aid in this process we computed periodograms using three methods: Lomb-Scargle \citep{lomb:1976, scargle:1982}, Analysis of Variance \citep{schwarzenberg:1989}, and Box-fitting Least Squares (BLS) \citep{kovacs:2002}, as implemented in the vartools package \citep{hartman:2012}. Systems with equal depth eclipses may be revised in the future with additional follow-up data (Section \ref{sec:spectra}).

\subsection{Period and Ephemeris Error Estimates}
\label{sec:EEE}

We now provide error estimates on both the period and time of eclipse ($\mathrm{BJD}_0$) for every eclipsing binary in the catalog. The period error is determined through an adaptation of the Period Error Calculator algorithm of \citet{mighell:2013}. Using error propagation theory, the period error is calculated from the following parameters: timing uncertainty for a measured flux value, the total length of the time series, the period of the variable, and the maximum number of periods that can occur in the time series. 

To revise a precise value and estimate the uncertainty on $\mathrm{BJD}_0$, we use the eclipse bisectors. The bisectors work by shifting the $\mathrm{BJD}_0$ until the left and the right eclipse sides of the phased data overlap as much as possible. The overlap function is fitted by a gaussian, where the mean is the $\mathrm{BJD}_0$ estimate, and the width is the corresponding error. For those in which this estimates an error larger than the measured width of the eclipse, generally due to extremely low signal to noise, we assume the width of the eclipse as the error instead.  Since there isn't a well defined $\mathrm{BJD}_0$ for binary heartbeat stars that do not exhibit an actual eclipse in the light curve, we do not estimate or provide uncertainties for these objects.

\section{Catalog Deletions}
\label{sec:falseEB}
In order to provide the Eclipsing Binary community with the most accurate information, we checked each of the \numberCatalogEBs{} systems against extrinsic variability (i.e. third light contamination, cross talk, and other instrumental artifacts). We generated diagnostic plots for each available quarter of data for every object in the \emph{Kepler} Eclipsing Binary catalog to confirm whether the target was the true source of the EB signal; if the target was not the source of the signal it was deemed a FP.

\begin{figure}[ht!]
\includegraphics[width=\hsize,height=8.5cm]{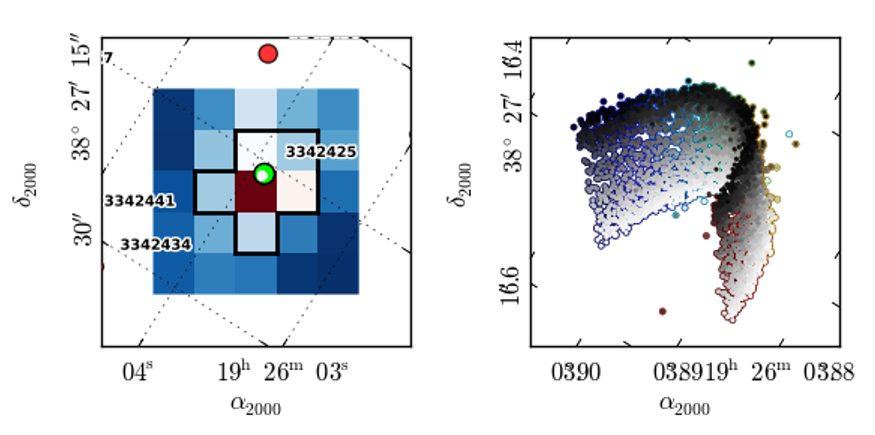}
\caption{\small Diagnostic plots for one quarter. In this example the target in question (KIC 3342425) is responsible for the binary signal. In the left-hand plot we see that the pixel contributing the most flux is under the target. On the right-hand side we observe the centroid movement in time. The dispersion in the y-direction (for this particular system) is due to the target eclipsing, causing the flux to decrease and the centroid to migrate towards the brighter star, which can be seen in the upper part of the left-hand plot. }
\label{fig:fpPlot1}
\end{figure}

\begin{figure}[ht!]
\includegraphics[width=\hsize,height=8.5cm]{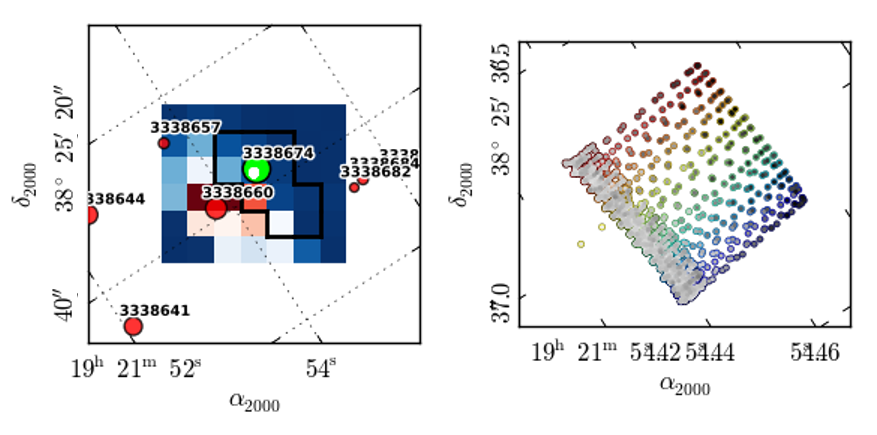}
\caption{\small Diagnostic plots for one quarter. In this example the target in question (KIC 3338674) is not responsible for the binary signal seen. On the left-hand side we see that the pixels contributing the most signal are not associated with our target. On the right-hand side we see that the centroid is constantly pulled elsewhere while only returning to our target when the off-target star is in eclipse. The binary light curve is generated by KIC 3338660, making KIC 3338674 a false positive.}
\label{fig:fpPlot2}
\end{figure}

To do this, we design and generate two diagnostic plots per quarter that - when considered together with all other data - give insight into which object in the target pixel file (TPF) map is responsible for the binary signal observed in the light curve. Figs. \ref{fig:fpPlot1} and \ref{fig:fpPlot2} show a heat map (left panels) that depicts the correlation of the eclipse depth of each individual pixel within the \emph{Kepler} aperture (outlined in black) compared to the eclipse depth of the summed light curves of all the pixels in the aperture. The color scale of the heat map is normalized to the summed \emph{Kepler} light curve divided by the number of pixels in the aperture; the average of pixels in the mask is assigned to white. The heat map shows where the signal contribution is located. Red colors represent pixels with a higher signal contribution while blue colors represent pixels with a lower signal contribution than the average-value pixels. A green circle represents the target KIC that is being processed while any other KIC objects in the frame are plotted with red circles. The coordinates of these circles come from KIC catalog positions. The radius of the circles correspond to the \emph{Kepler} magnitude. The collection of white dots, typically located near the green circle, represent the location and shape of the centroid. The centroid is the center of light in the pixel window at a given time during the quarter.

The second diagnostic plot (Fig. \ref{fig:fpPlot1} and \ref{fig:fpPlot2}, right panels) illustrates an expanded view of the centroid movement throughout a given quarter. The varying colors of each circle edge represent the transition in time from the beginning of the quarter (blue) to the end of the quarter (red). The gray-scale color of each circle represents the detrended flux of the light curve at that moment, with darker shades representing lower flux. As the binary eclipses, the flux decreases and the centroid migrates towards areas of higher flux (typically away from the binary source). This, in conjunction with the heat map, gives insight into which object is responsible for the binary activity. 

These diagnostic plots, in conjunction with other data, were analyzed and Eclipsing Binary Working Group members voted on which object in the window was responsible for the binary signal. This was a blind vote with the results tabulated continuously. If the decision was not unanimous the targets were discussed in open group sessions using additional resources from data validation (DV) disposition reports \citep{coughlin:2014} when available at NExScI\footnotemark.

\footnotetext{\texttt{http://nexsci.caltech.edu/}} 

In addition to checking extrinsic variability, ephemeris cross-matching was performed across the Catalog to identify more FPs. This cross-matching solved ambiguity between double and half-period systems (for systems whose primary and secondary shape are identical). Any matches found this way were manually inspected to identify the true source and remove the FP.

%A target could be voted on by any number of members; at least three votes per target are required for validation. Triage sessions were held to resolve cases where voting results were split...The results of the voting and triage sessions were then used to update the target status in the Catalog.

\section{Light Curve Deconvolution}
\label{sec:LCD}
\begin{figure}[ht!]
\hfill{}
\includegraphics[width=\hsize,height=7.5cm]{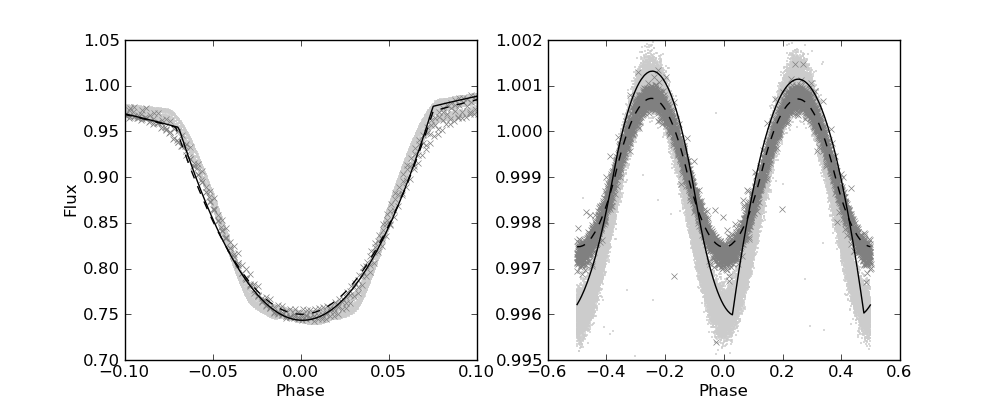}\\
\caption{\small Original polyfit (dashed line) and deconvolved polyfit (solid line) plotted on top of short-cadence (light dots) and long-cadence (darker x) Q2 data of KIC 11560447 and 6947064.  In both cases, the deconvolution was successful in finding a polyfit which when convolved best fits the long-cadence data, but this does not necessarily fit the short-cadence data or make physical sense.}
\label{fig:convolution}
\hfill{}
\end{figure}

Due to \emph{Kepler}'s long-cadence 30 minute exposure, the phased light curves suffer from a convolution effect.  For short-period binaries this phase-dependent smoothing can have a significant impact on the overall shape of the polyfit representation, which would then propagate through the remaining steps of the pipeline. To mitigate this effect, we deconvolve the original polyfit that was determined from the phased long-cadence data. Since there are an infinite number of functions that would convolve to the original polyfit, we impose that the deconvolved representation must also be described by a polyfit.  This does not necessarily guarantee a unique solution, but does add the constraint that the deconvolved curve resembles the signal from a binary. We start with the original polyfit and use a downhill simplex algorithm to adjust the various coefficients and knots, minimizing the residuals between the original polyfit and the convolved candidate-polyfit.  This process results in another polyfit that, when convolved with a 30 minute boxcar, most closely resembles the original polyfit and, therefore, the phased long-cadence data (Fig. \ref{fig:convolution}).  This results in a better representation of the actual light curve of the binary which can then be used to estimate geometrical properties of the binary.

\emph{Kepler}'s short-cadence photometry (1 min exposures) is short enough that the convolution effect is negligible. However, short cadence data are not available for all systems. Thus, to stay internally consistent, we use only long-cadence data to determine physical parameters and eclipse timing variations. Nevertheless, short-cadence data allow us to confirm the effect of convolution and test the performance of our deconvolution process.  Comparing the deconvolved polyfit with the short-cadence data from the same EB shows that deconvolution is essential for more accurate approximations but can also result in a representation that does not make physical sense (see the right panel in \ref{fig:convolution}). For this reason, if the deconvolved polyfit for a short-period EB looks suspicious, the deconvolution process likely introduced undesired artifacts into the light curve representation.

%We caution that any study of an individual EB should also take any available short-cadence data into account, and these data are also provided in the online Catalog.  

%\sout{\textbf{At a period of 100 hours}, the 30-minute boxcar in phase-space becomes the same width as the sampling of the polyfit.  \Since LLE assumes a deconvolved polyfit as input, we use the long-cadence deconvolved polyfit for any EB with a period less than 10 days \textbf{(well detached binaries are on the order of 10 days or more)}.  For longer periods, the convolution effect becomes insignificant at the resolution used in geometric analysis and we can safely use the original long-cadence polyfit.}

\section{Classification of Light Curves}
\label{sec:LLE}
EB light curves come in a variety of different shapes that are governed by a number of parameters. In order to overcome the drawbacks of manual classification, we performed an automated classification of all identified EB light curves with a general dimensionality reduction numerical tool called Locally Linear Embedding (LLE; \citealt{ roweis:2000}). This method is able to project a high-dimensional data-set onto a much lower-dimensional manifold in such a way that it retains the local properties of the original data-set, so all light curves that are placed close together in the original space are also nearby each other in the projected space (Paper III). This way the relations between the light curves are much easier to investigate. The input space is represented by a collection of polyfit models corresponding to each of the EB light curves in the Catalog. All polyfits were calculated in 1000 equidistant phase points (representing the original space) and then projected to a 2-dimensional space, resulting in an arc-shaped manifold. Inspection of the underlying light curves along the arc revealed that one end of the arc was populated by well-detached binaries while the other hosted the overcontact binaries and systems with ellipsoidal variations. Semi-detached systems were in the middle. We assigned a single number to each of the light curves based on where on the arc their projection is located to provide an easy-to-use single-number quantitative representation of the light curves. Values of the parameter range from 0 to 0.1 for well-detached systems, values below 0.5 predominantly belong to detached systems and between 0.5 and 0.7 to semi-detached systems. Overcontact systems usually have values between 0.7 and 0.8, while even higher values up to 1 usually belong to ellipsoidal variables.

In addition to the classification number, we also provide the depths and widths of the primary and secondary eclipses, as well as the separations between eclipses from the polyfit.  Depths are determined from the amplitude of the polynomial fitted to the eclipse and are in units of normalized flux.  A depth measurement is provided so long as it is larger than three times the estimated scatter of the light curve baseline.  For this reason, values are not provided for a secondary eclipse if the signal to noise is not sufficient.  Widths are determined by the ``knots'' connecting the individual polynomial sections and are in units of phase.  Separations are determined as the distance between the primary and secondary minimum, also in units of phase, and are defined to always be less than or equal to 0.5.  These values are only approximate measures and are only as accurate as the polyfits themselves.  Nevertheless, they do provide some measure of the signal-to-noise and also allow for an estimate of eccentricities (Pr\v sa et al. 2015).

\section{Visualizing Kepler data}
\label{sec:tSNE}

Visualization is an important aspect of data mining -- it allows a more intuitive and interactive approach to analyzing and interpreting the data-set. The employed algorithm, t-Distributed Stochastic Neighbor Embedding (t-SNE; \citealt{maaten:2008}) is particularly useful for high-dimensional data that lie on several different, but related, lower-dimensional manifolds. It allows the user to simultaneously view EB properties from multiple viewpoints. 

\subsection{The t-SNE algorithm}

An extension to the EB classification with LLE has been performed using a new technique of visualizing high-dimensional data, first proposed and developed by \cite{maaten:2008}. This technique, called t-SNE, is a modified version of the Stochastic Neighbor Embedding technique and has a specific appeal for visualizing data, since it is capable of revealing both global and local structure in terms of clustering data with respect to similarity. We provide a brief overview of the basic principles of the t-SNE technique, the results of its application to \emph{Kepler} data and its implementation in the interactive visualization of the Catalog.

Stochastic Neighbor Embedding defines data similarities in terms of conditional probabilities in the high-dimensional data space and their low dimensional projection. Neighbors of a data-point in the high dimensional data space are picked in proportion to their probability density under a Gaussian. Therefore, the similarity of two data-points is equivalent to a conditional probability. In t-SNE, the conditional probability is replaced by a joint probability that depends upon the number of data-points. This ensures that all data-points contribute to the cost function by a significant amount, including the outliers. The conditional probability of the corresponding low dimensional counterparts in SNE is also defined in terms of a Gaussian probability distribution, but t-SNE has introduced a symmetrized Student t-distribution, which leads to joint probabilities of the map. This allows for a higher dispersion of data-points in the low dimensional map and avoids unwanted attractive forces, since a moderate distance in the high-dimensional map can be represented well by larger distances in its low-dimensional counterpart. 

An input parameter that defines the configuration of the output map is the so-called perplexity. Therefore, instead of providing the desired number of nearest neighbors, the user provides a desired value of the perplexity and leaves it up to the method to determine the number of nearest neighbors, based on the data density. This in turn means that the data itself affects the number of nearest neighbors, which might vary from point to point.

\subsection{t-SNE and {\sl Kepler} data}

We have applied t-SNE to various samples of {\sl Kepler} data and compared the results to other methods (LLE, manual flags). We have also attempted two different approaches to the two-dimensional projection: one with direct high-dimensional to 2-dimensional mapping and one equivalent to the LLE classification method already implemented in the Catalog (Paper III), where the high-dimensional data space is first mapped to a 3D projection, and subsequently mapped to a 2D projection. The latter approach performs better in terms of clustering data-points and reveals a lumpier data structure, while the first approach results in a rather continuous structure, which might render a more elegant visualization of certain parameter distributions over the projection.

Both the two-step and the direct projection seem to be in agreement with the results obtained with LLE. One advantage of LLE over t-SNE is its simple structure, which allowed for a quantitative classification in terms of the morphology parameter (Paper III). The complex two-dimensional structure of the t-SNE projection makes that task substantially more difficult, so we use t-SNE only as a visualization tool, and retain LLE for classification.

\begin{figure}[h]
\begin{center}
\includegraphics[scale=0.4]{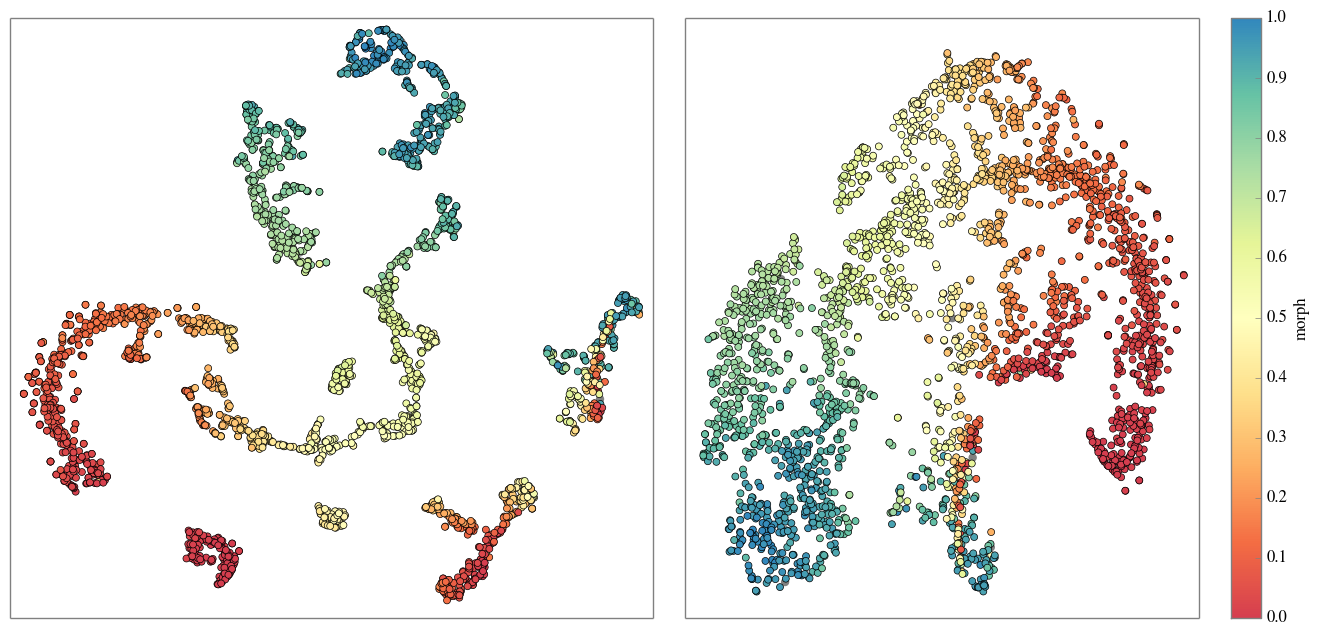}
\caption{\small Distribution of the LLE morphology parameter over the two-step (panel a) and direct 2D t-SNE projection (panel b).}
\label{tsnemorph}
\end{center}
\end{figure}

The distribution of the LLE morphology parameter on the t-SNE projection is depicted in Fig.~\ref{tsnemorph}. The concordance between the two methods is evident from the continuous variation of the morphology parameter along the map. It also serves to further illustrate the performance of t-SNE in terms of large-scale and small-scale structure. For example, the two islands in the right corner of the two-step projection (Fig \ref{tsnemorph}, panel (a)), which incorporate mainly noisy and unique light curve data, still manage to retain the large-scale gradient of the morphology parameter directed from the bottom to the top of the projection.

Further comparison also shows the two-step t-SNE ability to distinguish the main types of eclipsing binaries: objects with a morphology parameter $c$ (Paper III) $\lesssim 0.4$, corresponding to detached binaries, are a separate cluster from those with $0.4 \lesssim c \lesssim 0.8$ and those with $0.7 \lesssim c \lesssim 0.8$. Those with $c \gtrsim 0.8$, corresponding to ellipsoidal variations and objects with uncertain classification, are grouped into the topmost cluster, separate from all the previous ones. Based on this, it is safe to conclude that t-SNE also performs in line with the manual classification and it might be useful in speeding up or even replacing the subjective manual classification process.

\begin{figure}[h]
\begin{center}
\includegraphics[scale=0.5]{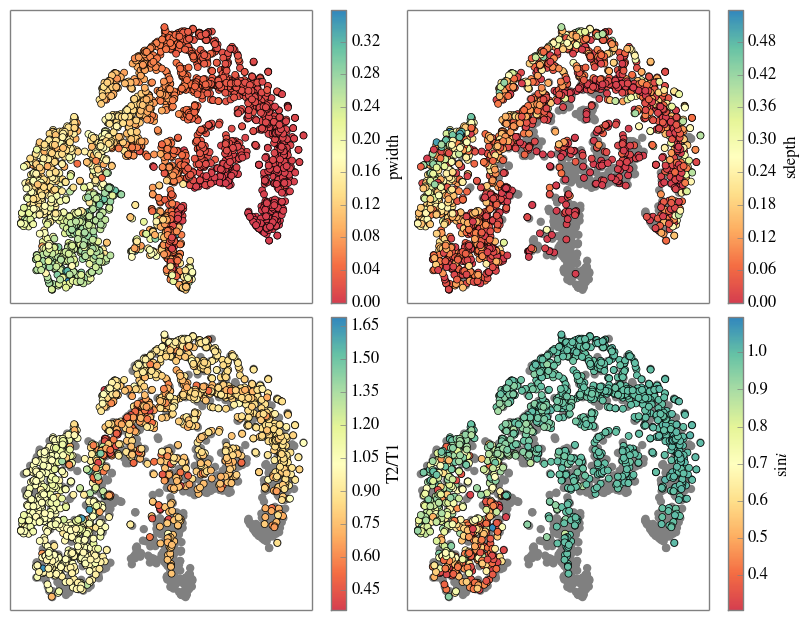}

\caption{\small Distribution of primary eclipse width (panel a), secondary eclipse depth (panel b), $T2/T1$ (panel c) and $\sin i$ (panel d) parameters over the direct 2D t-SNE projection. The data-points that don't have a value for the desired parameter are marked in grey.}
\label{tsneparams}
\end{center}
\end{figure}

To further illustrate the performance of the technique, we have generated plots of all cataloged parameters over the projection. In all cases the distribution seems fairly smooth, which attests to the broad range of technique applicability. Plots of the distribution of primary eclipse width and secondary eclipse depth obtained with polyfit, and the ratio of temperatures ($T2/T1$) and $\sin i$ obtained through geometric analysis, are provided in Fig.~\ref{tsneparams} (panel (a), (b), (c), and (d) respectively).

An example of the power of tSNE to reveal substructure and distinguish between many different types of similarities which may arise from parameters such as inclination, primary eclipse widths, secondary eclipse depth etc., within a given classification can be seen in the ``branches'' of Fig. \ref{tsnemorph} (panel b) and revealed in Fig. \ref{tsneparams}. The two groups of light curves corresponding to ellipsoidal variations ($c \gtrsim 0.8$) at the bottom of Fig. \ref{tsnemorph} (panel b) are especially indicative of this, since only the left branch corresponds to manually classified ellipsoidal light curves, while the right branch, although having a morph parameter that corresponds to ellipsoidal curves, is in fact composed of noisy or unique light curves that can not be classified into any of the primary morphological types. Most of the heartbeat stars can be found at the bottom of this branch.

To fully benefit from the capabilities of the technique, we implemented an interactive version in the Catalog, which combines the t-SNE visualization of {\sl Kepler} data with a broad range of parameter distributions. This extends the data  possibilities to a more intuitive and interactive approach, enabling the user to directly view the properties of certain light curves with respect to the whole Catalog. This can be found at \texttt{http://keplerEBs.villanova.edu/tsne}.

\section{Interesting Classes of Objects in the Catalog}

\subsection{Heartbeat Stars}
\label{sec:HB}

\begin{figure}[H]
\hfill{}%,angle=270
\center
\includegraphics[width=\linewidth]{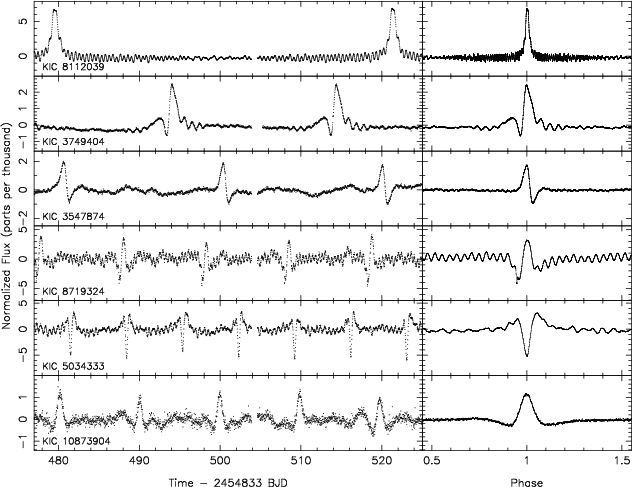}
\caption{\small Time series of a selection of \emph{Kepler} heartbeat star light curves (left panel) and phase folded data (right panel). The phase folded data clearly depict the oscillations that are integer multiples of the orbital frequency: tidally induced pulsations. It is worth noting that KIC 8112039 is KOI 54 \citep{welsh:2011}.}
\label{fig:HBs} 
\hfill{}
\end{figure}

Heartbeat stars are a subclass of eccentric ellipsoidal variables introduced by \citet{thompson:2012}. The most prominent feature in the heartbeat star light curve is the increased brightness at periastron (where periastron is defined as the closest approach of the two binary star components) caused by stellar deformation, which is a consequence of gravitational interactions; and heating (Fig. \ref{fig:HBs}). The morphology of the heartbeat-star light curve defines the heartbeat star -- a periastron variation preceded and succeeded by a flat region (ignoring pulsations and spots etc.). As shown by \citet{kumar:1995}, the shape of the brightening primarily depends on three orbital properties: argument of periastron, eccentricity, and inclination. This is true for the majority of objects where irradiation effects cause only minor modifications \citep{burkart:2012}. However, this effect is temperature dependent and so for objects with hotter components that both contribute significantly to the flux, as in the case of KOI-54 \citep{welsh:2011}, which contains two A stars, the irradiation contributes notably to the light curve (approximatly half of the amplitude of the periastron variation of KOI-54 is a consequence of irradiation). Overall, the amplitude of the periastron brightening depends on the temperature and structure of the stellar components, and the periastron distance. This enables heartbeat stars to be modeled without the presence of eclipses and thus at any inclination. Table\,\ref{tab:HB} contains the \emph{Kepler} catalog identifiers (KIC) and corresponding periods for 173 currently known heartbeart stars in the {\it Kepler} sample. These systems are flagged with the ``HB'' flag in the Catalog.

\begin{center}
\scriptsize \begin{longtable}{c c c c} 
\hline 
\hline 
KIC&\multicolumn{1}{p{2cm}}{\centering Period\\(days)}&\multicolumn{1}{p{2cm}}{\centering RA($^\circ$)\\(J2000)}&\multicolumn{1}{p{2cm}}{\centering Dec($^\circ$)\\(J2000)}\\ 
\hline
\endfirsthead

\multicolumn{4}{c}%
{\tablename\ \thetable\ -- \textit{Continued from previous page}}\\
\hline
\hline 
KIC&\multicolumn{1}{p{2cm}}{\centering Period\\(days)}&\multicolumn{1}{p{2cm}}{\centering RA($^\circ$)\\(J2000)}&\multicolumn{1}{p{2cm}}{\centering Dec($^\circ$)\\(J2000)}\\ 
\hline
\endhead

\hline \multicolumn{4}{r}{\textit{Continued on next page}} \\
\endfoot

\hline
\endlastfoot

1573836& 3.557093 & 291.5025&37.1775\\
2010607& 18.632296 & 290.5056&37.4590\\
2444348& 103.206989 & 291.6687&37.7041\\
2697935& 21.513359 & 287.4679&37.9666\\
2720096& 26.674680 & 292.9791&37.9109\\
3230227& 7.047106 & 290.1126&38.3999\\
3240976& 15.238869 & 292.9291&38.3280\\
3547874& 19.692172 & 292.2829&38.6657\\
3729724& 16.418755 & 285.6555&38.8505\\
3734660& 19.942137 & 287.6238&38.8554\\
3749404& 20.306385 & 292.0795&38.8371\\
3764714& 6.633276 & 295.7400&38.8623\\
3766353& 2.666965 & 296.0538&38.8943\\
3850086& 19.114247 & 291.3975&38.9647\\
3862171& 6.996461 & 294.4970&38.9801\\
3869825& 4.800656 & 296.1339&38.9990\\
3965556& 6.556770 & 294.3909&39.0767\\
4142768& 27.991603 & 287.2629&39.2600\\
4150136& 9.478402 & 289.6193&39.2939\\
4247092& 21.056416 & 286.8797&39.3784\\
4248941& 8.644598 & 287.6032&39.3977\\
4253860& 155.061112 & 289.2337&39.3865\\
4359851& 13.542328 & 290.0829&39.4008\\
4372379& 4.535171 & 293.5203&39.4478\\
4377638& 2.821875 & 294.7438&39.4994\\
4450976& 12.044869 & 287.3733&39.5261\\
4459068& 24.955995 & 290.0244&39.5634\\
4470124& 11.438984 & 293.1966&39.5072\\
4545729& 18.383520 & 286.4008&39.6984\\
4649305& 22.651138 & 290.1033&39.7816\\
4659476& 58.996374 & 293.1102&39.7564\\
4669402& 8.496468 & 295.5339&39.7623\\
4761060& 3.361391 & 295.5471&39.8618\\
4847343& 11.416917 & 294.9517&39.9294\\
4847369& 12.350014 & 294.9584&39.9046\\
4936180& 4.640922 & 295.1137&40.0712\\
4949187& 11.977392 & 297.9250&40.0884\\
4949194& 41.263202 & 297.9279&40.0549\\
5006817& 94.811969 & 290.4560&40.1457\\
5017127& 20.006404 & 293.5486&40.1117\\
5034333& 6.932280 & 297.4004&40.1495\\
5039392& 236.727941 & 298.4009&40.1722\\
5090937& 8.800693 & 289.2657&40.2555\\
5129777& 26.158530 & 299.0478&40.2189\\
5175668& 21.882115 & 288.3171&40.3280\\
5213466& 2.819311 & 298.1299&40.3999\\
5284262& 17.963312 & 294.4163&40.4802\\
5286221& 15.295983 & 294.9404&40.4966\\
5398002& 14.153175 & 300.0150&40.5270\\
5511076& 6.513199 & 282.9722&40.7270\\
5596440& 10.474857 & 282.7207&40.8992\\
5707897& 8.416091 & 292.5603&40.9969\\
5733154& 62.519903 & 298.6118&40.9491\\
5736537& 1.761529 & 299.2878&40.9129\\
5771961& 26.066437 & 284.7599&41.0434\\
5790807& 79.996246 & 291.7718&41.0981\\
5818706& 14.959941 & 298.6959&41.0417\\
5877364& 89.648538 & 292.1599&41.1982\\
5944240& 2.553222 & 285.9139&41.2916\\
5960989& 50.721534 & 292.0520&41.2661\\
6042191& 43.390923 & 291.7434&41.3100\\
6105491& 13.299638 & 284.9915&41.4333\\
6117415& 19.741625 & 289.8611&41.4082\\
6137885& 12.790099 & 295.8830&41.4747\\
6141791& 13.659035 & 296.7550&41.4846\\
6290740& 15.151827 & 293.6252&41.6615\\
6292925& 13.612220 & 294.2413&41.6198\\
6370558& 60.316584 & 293.8644&41.7169\\
6693555& 10.875075 & 292.4559&42.1372\\
6775034& 10.028547 & 291.0820&42.2686\\
6806632& 9.469157 & 298.8907&42.2099\\
6850665& 214.716056 & 287.7663&42.3227\\
6881709& 6.741116 & 296.6588&42.3698\\
6963171& 23.308219 & 295.8099&42.4592\\
7039026& 9.943929 & 293.4486&42.5866\\
7041856& 4.000669 & 294.2317&42.5901\\
7050060& 22.044000 & 296.2551&42.5302\\
7259722& 9.633226 & 283.1575&42.8904\\
7293054& 671.800000 & 295.0398&42.8795\\
7350038& 13.829942 & 287.5149&42.9118\\
7373255& 13.661106 & 294.8559&42.9372\\
7431665& 281.400000 & 287.3766&43.0095\\
7511416& 5.590855 & 286.0600&43.1662\\
7591456& 5.835751 & 285.6098&43.2055\\
7622059& 10.403262 & 295.9200&43.2818\\
7660607& 2.763401 & 281.8825&43.3002\\
7672068& 16.836177 & 287.8638&43.3042\\
7799540& 60.000000 & 281.1918&43.5250\\
7833144& 2.247734 & 295.3246&43.5054\\
7881722& 0.953289 & 288.2734&43.6977\\
7887124& 32.486427 & 290.4833&43.6224\\
7897952& 66.991639 & 294.2019&43.6553\\
7907688& 4.344837 & 297.0531&43.6438\\
7914906& 8.752907 & 298.8165&43.6733\\
7918217& 63.929799 & 299.6020&43.6960\\
7973970& 9.479933 & 296.3820&43.7489\\
8027591& 24.274432 & 291.3655&43.8680\\
8095275& 23.007350 & 290.9708&43.9708\\
8112039& 41.808235 & 296.5647&43.9476\\
8123430& 11.169990 & 299.3394&43.9996\\
8144355& 80.514104 & 281.8770&44.0299\\
8151107& 18.001308 & 285.5886&44.0477\\
8164262& 87.457170 & 291.2468&44.0004\\
8197368& 9.087917 & 300.9348&44.0943\\
8210370& 153.700000 & 280.7721&44.1887\\
8242350& 6.993556 & 294.9346&44.1298\\
8264510& 5.686759 & 300.9686&44.1920\\
8322564& 22.258846 & 298.7603&44.2662\\
8328376& 4.345967 & 300.3823&44.2801\\
8386982& 72.259590 & 298.4551&44.3134\\
8456774& 2.886340 & 299.7266&44.4091\\
8456998& 7.531511 & 299.7784&44.4611\\
8459354& 53.557318 & 300.4068&44.4142\\
8508485& 12.595796 & 296.4474&44.5586\\
8688110& 374.546000 & 291.3925&44.8952\\
8696442& 12.360553 & 294.5535&44.8503\\
8702921& 19.384383 & 296.6650&44.8531\\
8703887& 14.170980 & 296.9350&44.8503\\
8707639& 7.785196 & 297.8877&44.8683\\
8719324& 10.232698 & 301.1143&44.8258\\
8803882& 89.630216 & 284.7969&45.0991\\
8838070& 43.362724 & 298.1911&45.0936\\
8908102& 5.414582 & 298.8360&45.1165\\
8912308& 20.174443 & 300.0684&45.1016\\
9016693& 26.368027 & 289.8839&45.3041\\
9151763& 438.051939 & 290.6850&45.5684\\
9163796& 121.006844 & 295.3375&45.5048\\
9408183& 49.683544 & 293.4932&45.9876\\
9535080& 49.645296 & 295.2635&46.1480\\
9540226& 175.458827 & 297.0340&46.1985\\
9596037& 33.355613 & 294.9424&46.2501\\
9701423& 8.607397 & 287.6039&46.4128\\
9711769& 12.935909 & 292.4648&46.4918\\
9717958& 67.995354 & 294.9158&46.4853\\
9790355& 14.565548 & 298.7419&46.5773\\
9835416& 4.036605 & 293.6965&46.6205\\
9899216& 10.915849 & 295.1617&46.7506\\
9965691& 15.683195 & 297.3268&46.8452\\
9972385& 58.422113 & 299.2823&46.8988\\
10004546& 19.356914 & 288.7470&46.9307\\
10092506& 31.041675 & 297.8457&47.0723\\
10096019& 6.867469 & 298.8533&47.0828\\
10159014& 8.777397 & 297.7946&47.1565\\
10162999& 3.429215 & 298.8757&47.1436\\
10221886& 8.316718 & 296.8986&47.2816\\
10334122& 37.952857 & 289.6644&47.4004\\
10611450& 11.652748 & 296.1244&47.8307\\
10614012& 132.167312 & 296.9287&47.8830\\
10664416& 25.322174 & 291.4008&47.9246\\
10679505& 5.675186 & 296.9342&47.9812\\
10863286& 3.723867 & 292.7082&48.2070\\
10873904& 9.885633 & 296.6001&48.2960\\
11044668& 139.450000 & 297.9449&48.5578\\
11071278& 55.885225 & 284.4661&48.6553\\
11122789& 3.238154 & 283.1246&48.7349\\
11133313& 27.400893 & 289.6666&48.7296\\
11240948& 3.401937 & 290.1068&48.9034\\
11288684& 22.210063 & 287.1591&49.0800\\
11403032& 7.631634 & 291.9279&49.2970\\
11409673& 12.317869 & 295.1364&49.2733\\
11494130& 18.955414 & 284.0268&49.4154\\
11506938& 22.574780 & 291.6465&49.4934\\
11568428& 1.710629 & 296.1853&49.5568\\
11568657& 13.476046 & 296.2798&49.5178\\
11572363& 19.027753 & 297.7151&49.5738\\
11649962& 10.562737 & 284.5680&49.7795\\
11700133& 6.754017 & 284.2339&49.8532\\
11769801& 29.708220 & 295.1429&49.9931\\
11774013& 3.756248 & 296.8993&49.9163\\
11859811& 22.314148 & 289.3813&50.1894\\
11923629& 17.973284 & 296.3860&50.2050\\
11970288& 20.702319 & 295.2126&50.3169\\
12255108& 9.131526 & 289.5991&50.9386\\

\hline 
\caption{The heartbeat stars in the {\it Kepler} sample.}
\label{tab:HB}
\normalsize 
\end{longtable}
\end{center}

\begin{figure}[H]
\hfill{}
\includegraphics[width=\hsize, height=10cm]{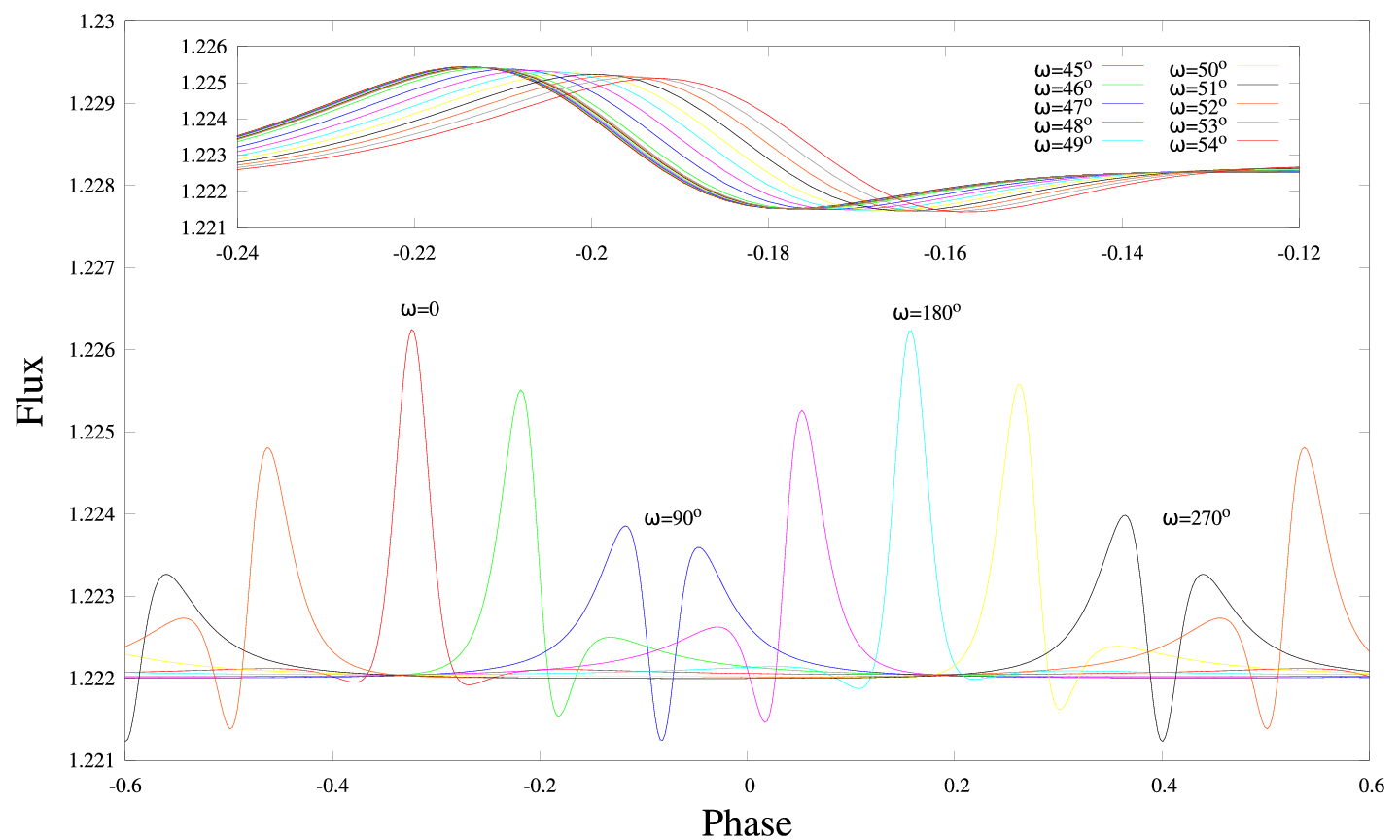}
\caption{\small The effect of the periastron value, $\omega$, on the shape of the periastron brightening for one full cycle. The peaks change both their position in phase and their shape drastically with varying $\omega$. KIC\,3749404 is a heartbeat star that exhibits one of the largest periastron advance rates ($\sim$2$^{\circ}$/yr; Hambleton et al., in prep). In the subplot the effect on the light curve over the four year {\it Kepler} mission is depicted.
\label{fig:Apsidal}
}
\hfill{}
\end{figure}

As the stellar components are in close proximity at periastron (relative to their radii), they are subject to strongly varying gravitational forces. Consequently, the stars experience large torques during periastron flybys, making them likely candidates for apsidal motion: the precession of the line of apsides about the center of mass \citep{claret:1993}. Since apsidal motion changes the argument of periastron, the advance can be determined by the change in the shape of the heartbeat light curve, assuming that the data-set is long enough for the change to be detected. Fig.~\ref{fig:Apsidal} shows an example of how the light curve of a heartbeat star changes as a function of the argument of periastron for the case of KIC\,3749404, which exhibits one of the largest periastron advance rates of $\sim$2$^{\circ}$/yr. The central density of the stellar components can be empirically inferred through the rate of the apsidal advance \citep{claret:2010}, enabling us to test our current understanding of stellar structure and evolution. Furthermore, as approximately half of all the known \emph{Kepler} heartbeat stars have periods $3<P<15\,d$ and high eccentricities ($e>0.3$), heartbeat stars are ideal for testing theories on relativistic apsidal motion \citep{gimenez:1985}.

\subsubsection{Tidally Induced Pulsations}

Approximately 15\% of these interesting objects also demonstrate tidally induced pulsations e.g. HD174884
\citep{maceroni:2009}, HD177863 \citep{willems:2002}, KOI-54 \citep{welsh:2011, fuller:2012, burkart:2012} and
KIC\,4544587 (\citealt{hambleton:2013}). The signature of a tidally induced pulsation is a mode (or modes) that is an integer multiple of the orbital frequency. From our sample, the record holder, KIC\,8164262, has a tidally induced pulsation amplitude of 1\,mmag. The majority of objects in our sample have tidally induced pulsations with maximum amplitudes of $\sim$0.1\,mmag. Tidally induced modes occur when the orbital frequency is close to an eigenfrequency of the stellar component, causing the star to act like a forced oscillator, which significantly increases the amplitude of the mode. \citet{zahn:1975} hypothesized the existence of tidally induced pulsations as a mechanism for the circularization of binary orbits, attributed to the exchange of orbital angular momentum and loss of orbital energy through mode damping. The high occurrence rate of tidally induced pulsations in heartbeat stars is due to their eccentric nature and small periastron distances leading to interaction times that are comparable to the eigenmodes of the stellar components. The right panel of Fig.\,\ref{fig:HBs} shows a selection of heartbeat star light curves folded on their orbital periods. The stellar pulsations are clearly visible as they are exact integer multiples of the orbital frequency: the signature of tidally induced pulsations. These 24 systems are flagged with the “TP” flag.

\begin{table}[ht!]
\center{}
\begin{center} 
\scriptsize \begin{tabular}{c c c c} 
\hline 
\hline 
KIC&\multicolumn{1}{p{2cm}}{\centering Period\\(days)}&\multicolumn{1}{p{2cm}}{\centering RA($^\circ$)\\(J2000)}&\multicolumn{1}{p{2cm}}{\centering Dec($^\circ$)\\(J2000)}\\ 
\hline
3230227& 7.047106  &290.1126&38.3999\\
3547874& 19.692172  &292.2829&38.6657\\
3749404& 20.306385  &292.0795&38.8371\\
3766353& 2.666965  &296.0538&38.8943\\
3869825& 4.800656  &296.1339&38.9990\\
4142768& 27.991603  &287.2629&39.2600\\
4248941& 8.644598  &287.6032&39.3977\\
4949194& 41.263202  &297.9279&40.0549\\
5034333& 6.932280  &297.4004&40.1495\\
5090937& 8.800693  &289.2657&40.2555\\
8095275& 23.007350  &290.9708&43.9708\\
8112039& 41.808235  &296.5647&43.9476\\
8164262& 87.457170  &291.2468&44.0004\\
8264510& 5.686759  &300.9686&44.1920\\
8456774& 2.886340  &299.7266&44.4091\\
8703887& 14.170980  &296.9350&44.8503\\
8719324& 10.232698  &301.1143&44.8258\\
9016693& 26.368027  &289.8839&45.3041\\
9835416& 4.036605  &293.6965&46.6205\\
9899216& 10.915849  &295.1617&46.7506\\
11122789& 3.238154  &283.1246&48.7349\\
11403032& 7.631634  &291.9279&49.2970\\
11409673& 12.317869  &295.1364&49.2733\\
11494130& 18.955414  &284.0268&49.4154\\
 
\hline 
\end{tabular}
\caption{The systems with Tidally Induced Pulsations in the {\it Kepler} sample.}
\label{tab:TP}
\normalsize 
\end{center} 
\end{table}

\subsection{Reflection Effect Binaries}
\label{sec:REF}
The Reflection Effect is the mutual irradiation of the facing hemispheres of two stars in the binary system. This irradiation alters the temperature structure in the atmosphere of the star, resulting in an increased intensity and flux. This effect reveals itself with an increased flux level on the ingress and egress of the eclipse in the light curve. There are currently 36 of these systems (Table \ref{tab:REF}). These systems are flagged with the ``REF'' flag.

\begin{table}[ht!]
\center{}
\begin{center} 
\scriptsize \begin{tabular}{c c c c c c c} 
\hline 
\hline 
KIC&\multicolumn{1}{p{2cm}}{\centering Period\\(days)}&\multicolumn{1}{p{2cm}}{\centering Period Error\\(days)}&\multicolumn{1}{p{2cm}}{\centering BJD$_0$ \\ (-2400000)}&\multicolumn{1}{p{2cm}}{\centering BJD$_0$ Error\\(days)}&\multicolumn{1}{p{2cm}}{\centering RA($^\circ$)\\(J2000)}&\multicolumn{1}{p{2cm}}{\centering Dec($^\circ$) \\(J2000)}\\ 
\hline
2708156& 1.891272 & 0.000003 & 54954.336 & 0.046 & 290.2871&37.9365\\
3339563& 0.841232 & 0.000001 & 54965.378 & 0.051 & 290.7460&38.4431\\
3431321& 1.015049 & 0.000001 & 54954.418 & 0.066 & 287.7745&38.5040\\
3547091& 3.30558 & 0.00001 & 55096.3221 & 0.0177 & 292.0785&38.6315\\
4350454& 0.965658 & 0.000001 & 54964.943 & 0.224 & 287.0525&39.4879\\
4458989& 0.529854 & 0.000000 & 54954.1586 & 0.0337 & 290.0015&39.5339\\
5034333& 6.93228 & 0.00002 & 54954.028 & 0.054 & 297.4004&40.1495\\
5098444& 26.9490 & 0.0001 & 54984.023 & 0.082 & 291.5633&40.2678\\
5213466& 2.81931 & 0.00001 & 55165.534 & \nodata & 298.1299&40.3999\\
5736537& 1.76153 & 0.00001 & 54965.955 & \nodata & 299.2878&40.9129\\
5792093& 0.600588 & 0.000000 & 54964.6434 & 0.0331 & 292.2002&41.0137\\
6262882& 0.996501 & 0.000001 & 54965.112 & 0.064 & 282.6457&41.6087\\
6387887& 0.216900 & 0.000001 & 54999.9686 & 0.0056 & 298.1854&41.7340\\
6791604& 0.528806 & 0.000000 & 54964.6825 & 0.0284 & 295.6755&42.2406\\
7660607& 2.76340 & 0.00001 & 54954.589 & \nodata & 281.8825&43.3002\\
7748113& 1.734663 & 0.000002 & 54954.144 & 0.102 & 290.1135&43.4365\\
7770471& 1.15780 & 0.00000 & 55000.1826 & 0.0263 & 297.1805&43.4771\\
7833144& 2.24773 & 0.00001 & 55001.232 & \nodata & 295.3246&43.5054\\
7881722& 0.953289 & 0.000002 & 54954.118 & \nodata & 288.2734&43.6977\\
7884842& 1.314548 & 0.000002 & 54954.876 & 0.082 & 289.6946&43.6260\\
8455359& 2.9637 & 0.0002 & 55002.027 & 0.189 & 299.3618&44.4112\\
8758161& 0.998218 & 0.000001 & 54953.8326 & 0.0243 & 293.6205&44.9673\\
9016693& 26.3680 & 0.0001 & 55002.583 & \nodata & 289.8839&45.3041\\
9071373& 0.4217690 & 0.0000003 & 54953.991 & 0.109 & 283.1591&45.4156\\
9101279& 1.811461 & 0.000003 & 54965.932 & 0.047 & 296.0016&45.4481\\
9108058& 2.1749 & 0.0001 & 54999.729 & 0.049 & 297.7811&45.4243\\
9108579& 1.169628 & 0.000001 & 54954.998 & 0.069 & 297.8999&45.4666\\
9159301& 3.04477 & 0.00001 & 54956.303 & 0.063 & 293.6933&45.5170\\
9472174& 0.1257653 & 0.0000001 & 54953.6432 & 0.0183 & 294.6359&46.0664\\
9602595& 3.55651 & 0.00001 & 54955.853 & 0.074 & 297.1435&46.2285\\
10000490& 1.400991 & 0.000002 & 54954.2437 & 0.0286 & 286.5560&46.9573\\
10149845& 4.05636 & 0.00001 & 55001.071 & 0.237 & 295.0444&47.1964\\
10857342& 2.41593 & 0.00003 & 55005.265 & 0.056 & 290.0115&48.2442\\
11408810& 0.749287 & 0.000001 & 54953.869 & 0.047 & 294.7304&49.2944\\
12109845& 0.865959 & 0.000003 & 55000.5582 & 0.0129 & 290.9192&50.6964\\
12216706& 1.47106 & 0.00001 & 55003.5216 & 0.0252 & 295.7438&50.8309\\

\hline 
\end{tabular}
\caption{The Reflection Effect systems in the {\it Kepler} sample. Those reported without $\mathrm{BJD}_0$ errors are also heartbeat stars.}
\label{tab:REF}
\normalsize 
\end{center} 
\end{table}

\subsection{Occultation Pairs}
\label{sec:OCC}

The majority of the stars that make up the Milky Way galaxy are M-type stars and, since they are faint because of their low mass, luminosity, and temperature, the only direct way to measure their masses and radii is by analyzing their light curves in EBs. An issue in the theory of these low-mass stars is the discrepancy between predicted and observed radii of M-type stars \citep{feiden2015}. When these stars are found in binaries with an earlier type component, the ratio in radii is quite large, and eclipses are total and occulting. Other extreme radius ratio pairs produce similar light curves, such as white dwarfs and main sequence stars, or main sequence stars and giants. Irrespective of the underlying morphology, these occultation pairs are critical gauges because eclipses are total and the models are thus additionally constrained.

There are currently 32 of these systems (Table \ref{tab:OCC}). These systems are flagged with the ``OCC'' flag.

\begin{table}[ht!]
\center{}
\begin{center} 
\scriptsize \begin{tabular}{c c c c c c c} 
\hline 
\hline 
KIC&\multicolumn{1}{p{2cm}}{\centering Period\\(days)}&\multicolumn{1}{p{2cm}}{\centering Period Error\\(days)}&\multicolumn{1}{p{2cm}}{\centering BJD$_0$ \\ (-2400000)}&\multicolumn{1}{p{2cm}}{\centering BJD$_0$ Error\\(days)}&\multicolumn{1}{p{2cm}}{\centering RA($^\circ$)\\(J2000)}&\multicolumn{1}{p{2cm}}{\centering Dec($^\circ$)\\(J2000)}\\ 
\hline
2445134& 8.4120089 & 0.0000228 & 54972.648 & 0.035 & 291.8497&37.7386\\
3970233& 8.254914 & 0.000055 & 54966.160 & 0.052 & 295.4819&39.0173\\
4049124& 4.8044707 & 0.0000102 & 54969.0039 & 0.0229 & 289.3142&39.1314\\
4386047& 2.9006900 & 0.0000181 & 55001.2005 & 0.0253 & 296.4303&39.4435\\
4740676& 3.4542411 & 0.0000063 & 54954.3478 & 0.0325 & 289.9644&39.8114\\
4851464& 5.5482571 & 0.0000159 & 55005.061 & 0.046 & 295.8887&39.9113\\
5370302& 3.9043269 & 0.0000076 & 54967.3507 & 0.0286 & 294.0568&40.5497\\
5372966& 9.2863571 & 0.0000262 & 54967.6753 & 0.0264 & 294.7440&40.5339\\
5728283& 6.1982793 & 0.0000153 & 55003.981 & 0.057 & 297.6203&40.9431\\
6182019& 3.6649654 & 0.0000074 & 55003.8010 & 0.0264 & 282.6849&41.5880\\
6362386& 4.5924016 & 0.0000094 & 54956.954 & 0.036 & 291.2642&41.7488\\
6387450& 3.6613261 & 0.0000069 & 54968.2691 & 0.0182 & 298.0917&41.7941\\
6694186& 5.5542237 & 0.0000124 & 54957.0531 & 0.0325 & 292.6311&42.1808\\
6762829& 18.795266 & 0.000072 & 54971.668 & 0.079 & 286.8304&42.2792\\
7037540& 14.405911 & 0.000063 & 55294.236 & 0.044 & 293.0204&42.5088\\
7972785& 7.3007334 & 0.0000187 & 54966.566 & 0.048 & 296.0477&43.7621\\
8230809& 4.0783467 & 0.0000081 & 54973.317 & 0.049 & 290.8917&44.1689\\
8458207& 3.5301622 & 0.0000065 & 54967.7749 & 0.0309 & 300.0916&44.4148\\
8460600& 6.3520872 & 0.0000153 & 54967.2065 & 0.0286 & 300.7748&44.4878\\
8580438& 6.4960325 & 0.0000158 & 54968.468 & 0.037 & 298.3827&44.6156\\
9048145& 8.6678260 & 0.0000238 & 54970.3408 & 0.0331 & 299.3980&45.3182\\
9446824& 4.2023346 & 0.0000087 & 55004.1071 & 0.0280 & 282.1844&46.0069\\
9451127& 5.1174021 & 0.0000111 & 54967.5858 & 0.0287 & 284.8227&46.0137\\
9649222& 5.9186193 & 0.0000138 & 54965.9239 & 0.0285 & 291.8585&46.3865\\
9719636& 3.351570 & 0.000044 & 55000.7919 & 0.0223 & 295.5262&46.4571\\
10020423& 7.4483776 & 0.0000192 & 54970.6936 & 0.0289 & 295.2979&46.9205\\
10295951& 6.8108248 & 0.0000167 & 54955.424 & 0.062 & 299.1028&47.3778\\
10710755& 4.8166114 & 0.0000102 & 54966.7629 & 0.0271 & 281.9364&48.0101\\
11200773& 2.4895516 & 0.0000039 & 54965.0434 & 0.0184 & 296.4774&48.8537\\
11252617& 4.4781198 & 0.0000096 & 55006.1732 & 0.0218 & 295.6914&48.9038\\
11404644& 5.9025999 & 0.0000137 & 54969.388 & 0.042 & 292.7561&49.2646\\
11826400& 5.8893723 & 0.0000135 & 54956.000 & 0.037 & 297.5134&50.0748\\
 
\hline 
\end{tabular}
\caption{The Occultation Pairs in the {\it Kepler} sample.}
\label{tab:OCC}
\normalsize 
\end{center} 
\end{table}

\subsection{Circumbinary Planets}
\label{sec:CBP}
The search for circumbinary planets in the \emph{Kepler} data includes looking for transits with multiple features \citep{deeg:1998,  doyle:2000}. Transit patterns with multiple features are caused by a slowly moving object crossing in front of the eclipsing binary; it is alternately silhouetted by the motion of the background binary stars as they orbit about each other producing predictable but non-periodic features of various shapes and depths. Emerging trends from studying these circumbinary planet systems show the planet’s orbital plane very close to the plane of the binary (in prograde motion) in addition to the host star. The orbits of the planets are in close proximity to the critical radius and the planet's sizes (in mass and/or radius) are smaller than that of Jupiter. A full discussion of these trends can be found in \cite{welsh:2014}. There are currently 14 of these systems (Table \ref{tab:CBP}). These systems are flagged with the ``CBP'' flag.

\begin{table}[H]

\center{}
\begin{center} 
\scriptsize \begin{tabular}{c c c c c c c c} 
\hline 
\hline 
KIC
&Kepler \#
&\multicolumn{1}{p{2cm}}{\centering Period\\(days)}
&\multicolumn{1}{p{2cm}}{\centering Period Error\\(days)}
&\multicolumn{1}{p{2cm}}{\centering RA($^\circ$)\\(J2000)}
&\multicolumn{1}{p{2cm}}{\centering Dec($^\circ$)\\(J2000)} 
& Citation\\
\hline
12644769& Kepler-16b &41.0776 & 0.0002 &289.075700&51.757400& \cite{doyle:2011}\\
8572936& Kepler-34b &27.7958 & 0.0001  &296.435800&44.641600& \cite{welsh:2012}\\
9837578& Kepler-35b &20.7337 & 0.0001  &294.497000&46.689800& \cite{welsh:2012}\\
6762829& Kepler-38b&18.7953 & 0.0001  &286.830400&42.279200& \cite{orosz:2012apj}\\
10020423& Kepler-47b,c &7.44838 & 0.00002  &295.297900&46.920500& \cite{orosz:2012sci}\\
4862625& Kepler-64b&20.0002 & 0.0001 &298.215100&39.955100& \cite{schwamb:2013}\\
12351927& Kepler413b&10.11615 & 0.00003  &288.510600&51.162500& \cite{kostov:2014}\\
9632895& Kepler-453b &27.3220 & 0.0001  &283.241300&46.378500& \cite{welsh:2015}\\

\hline 
\end{tabular}
\caption{\label{tab:CBP}The circumbinary planets in the {\it Kepler} sample.}
\normalsize 
\end{center} 
\end{table}

\subsection{Targets with Multiple Ephemerides}
Sources with additional features (\ref{sec:CBP}) can be another sign of a stellar triple or multiple system (in addition to ETV signals described in \citealt{orosz:2015} \& Orosz et al. 2015, in preparation). In this case the depth of the event is too deep to be the transit of a planet but is instead an eclipse by, or occultation of a third stellar body. A well known example is described in \cite{carter:2011}. We have been looking for such features in the Catalog and have uncovered 14 systems exhibiting multiple, determinable periods (Table \ref{tab:M}). These systems are flagged with the ``M'' flag in the Catalog. In some systems, extraneous events are observed whose ephemerides cannot be determined. In some cases the period is longer than the time baseline and two subsequent events have not been observed by \emph{Kepler}.  In other cases, eclipsing the inner-binary at different phases results in a non-linear ephemeris with an indeterminable period.  It is worth noting that without spectroscopy or ETVs that are in agreement that additional eclipse event is indeed related, these cases are not guaranteed to be multiple objects - some could be the blend of two independent binaries on the same pixel.  These events and their properties are reported in Table \ref{tab:EE}.

%(Fig. \ref{fig:TM}) 

%14 systems show up in the catalog results but that's because of redundancy of one of the systems due to the online catalog. This is due to individual ephemerides inheriting the M flag from the system. In the database this is correct with one M flag but the online system doesn't have that functionality.

\begin{table}[ht!]
\center{}
\begin{center} 
\scriptsize \begin{tabular}{c c c c c c c} 
\hline 
\hline 
KIC&\multicolumn{1}{p{2cm}}{\centering Period\\(days)}&\multicolumn{1}{p{2cm}}{\centering Period Error\\(days)}&\multicolumn{1}{p{2cm}}{\centering BJD$_0$ \\ (-2400000)}&\multicolumn{1}{p{2cm}}{\centering BJD$_0$ Error\\(days)}&\multicolumn{1}{p{2cm}}{\centering RA($^\circ$)\\ (J2000)}&\multicolumn{1}{p{2cm}}{\centering Dec($^\circ$) \\ (J2000)}\\ 
\hline
2856960& 0.2585073 & 0.0000001 & 54964.658506&0.007310&292.3813&38.0767\\
2856960& 204.256 & 0.002 & 54997.652563&0.369952&292.3813&38.0767\\
4150611& 8.65309 & 0.00002 & 54961.005419&0.024746&289.7425&39.2671\\
4150611& 1.522279 & 0.000002 & 54999.688801&0.003464&289.7425&39.2671\\
4150611& 94.198 & 0.001 & 55029.333888&0.328165&289.7425&39.2671\\
5255552& 32.4486 & 0.0002 & 54970.636491&0.116220&284.6931&40.4986\\
5897826& 33.8042 & 0.0002 & 54967.628858&0.116761&297.4759&41.1143\\
5952403& 0.905678 & 0.000001 & 54965.197892&0.014736&289.2874&41.2648\\
6665064& 0.69837 & 0.00001 & 54964.697452&0.009707&281.6500&42.1321\\
6964043& 5.36258 & 0.00002 & 55292.008176&0.308696&296.0145&42.4223\\
7289157& 5.26581 & 0.00001 & 54969.976049&0.044130&293.9661&42.8373\\
7289157& 242.713 & 0.002 & 54996.317389&0.055294&293.9661&42.8373\\
9007918& 1.387207 & 0.000002 & 54954.746682&0.023254&286.0084&45.3560\\
11495766& 8.34044 & 0.00002 & 55009.377729&0.046025&285.2253&49.4242\\
 
\hline 
\end{tabular}
\caption{The systems exhibiting Multiple ephemerides in the {\it Kepler} sample.}
\label{tab:M}
\normalsize 
\end{center} 
\end{table}

\begin{table}[ht!]
\center{}
\begin{center} 
\scriptsize \begin{tabular}{c c c c c} 
\hline 
\hline 
KIC
&\multicolumn{1}{p{2cm}}{\centering Event Depth\\ (\%)}
&\multicolumn{1}{p{2cm}}{\centering Event Width \\ (days)}
&\multicolumn{1}{p{2cm}}{\centering Start Time \\ (-240000)}
&\multicolumn{1}{p{2cm}}{\centering End Time\\ (240000)}\\
\hline
6543674& 0.96&2.&55023&55025\\
7222362& 0.6&0.6&55280.9&55281.5\\
7222362& 0.8&2&55307.5&55309.5\\
7222362& 0.65&2&55975.5&55977.5\\
7668648& 0.94&0.2&55501.1&55501.3\\
7668648& 0.94&0.2&55905.5&55905.7\\
7668648& 0.94&0.2&56104.4&56104.6\\
7668648& 0.94&0.2&56303.7&56303.9\\
7670485& 0.975&1.0&55663&55664\\

\hline 
\end{tabular}
\caption{Properties of the extraneous events found in the {\it Kepler} sample.}
\label{tab:EE}
\normalsize 
\end{center} 
\end{table}

\subsection{Eclipse Timing Variations}
\label{sec:ETV}
In an EB, one normally expects the primary eclipses to be uniformly spaced in time. However, apsidal motion, mass transfer from one star
to the other, or the presence of a third body in the system can give rise to changes in the orbital period, which in turn will
change the time interval between consecutive eclipse events. These deviations will contain important clues as to the reason for the period change. A table of these systems and more information on their analysis can be found in \citet{conroy:2014} and Orosz et al. (2015), in preparation. ETV values and plots are provided for each system in the online Catalog.

\subsection{Eclipse Depth Changes}
\begin{figure}[H]
\hfill{}
\center{}
\includegraphics[width=\hsize,height=12cm]{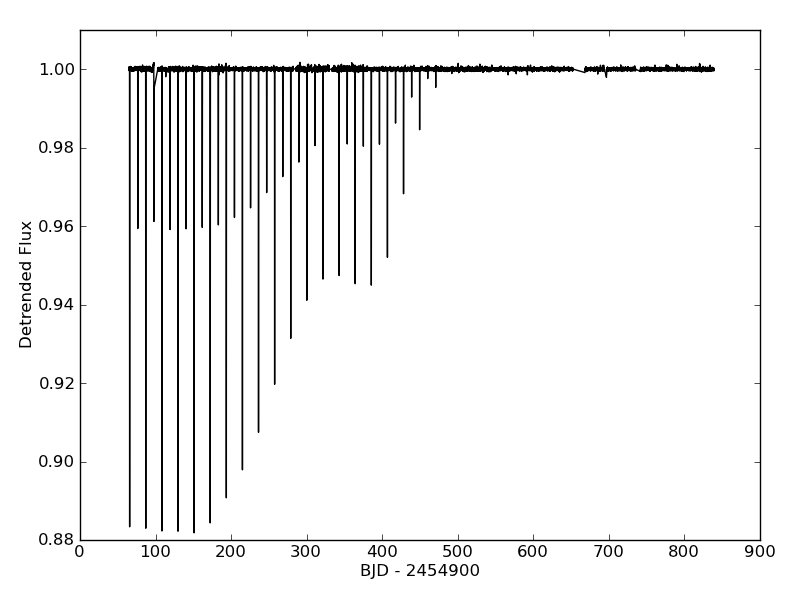}\\
\label{fig:10319590}
\caption{\small KIC 10319590 is a system undergoing eclipse depth variations.}
\label{fig:DV}
\hfill{}
\end{figure}

We have come across 43 systems that exhibit eclipse depth changes (Fig. \ref{fig:DV}). These systems were visually inspected and manually flagged. The depth variations shown here differ from QAM effects; the depth variations are long-term trends spanning multiple, sequential quarters. In 10319590 (see Fig.~\ref{fig:DV}), for example, the eclipses actually disappear. Although it is possible that some of these could be caused by gradual aperture movement or a source near the edge of a module leaking light, these variations are not quarter or season-dependent, and are much more likely to actually be physical.  Physical causes of these long-term depth variations could include a rapid inclination or periastron change due to the presence of a third body or spot activity as shown in \citealt{croll:2015}). Table \ref{tab:DV} lists these systems and are flagged by the ``DV'' (Depth Variation) flag.

\begin{table}[ht!]
\center{}
\begin{center}
\scriptsize \begin{tabular}{c c c c c c c} 
\hline 
\hline 
KIC&\multicolumn{1}{p{2cm}}{\centering Period\\(days)}&\multicolumn{1}{p{2cm}}{\centering Period Error\\(days)}&\multicolumn{1}{p{2cm}}{\centering BJD$_0$ \\ (-2400000)}&\multicolumn{1}{p{2cm}}{\centering BJD$_0$ Error\\(days)}&\multicolumn{1}{p{2cm}}{\centering RA($^\circ$)\\(J2000)}&\multicolumn{1}{p{2cm}}{\centering Dec($^\circ$)\\(J2000)}\\ 
\hline
1722276& 569.95 & 0.01 & 55081.923700&0.035212&291.6967&37.2380\\
2697935& 21.5134 & 0.0001 & 55008.469440&\nodata&287.4679&37.9666\\
2708156& 1.891272 & 0.000003 & 54954.335595&0.046299&290.2871&37.9365\\
3247294& 67.419 & 0.000 & 54966.433454&0.046494&294.4399&38.3808\\
3867593& 73.332 & 0.000 & 55042.964810&0.063167&295.6975&38.9020\\
3936357& 0.3691536 & 0.0000002 & 54953.852697&0.020054&285.4109&39.0412\\
4069063& 0.504296 & 0.000000 & 54964.906342&0.010072&294.5585&39.1377\\
4769799& 21.9293 & 0.0001 & 54968.505532&0.061744&297.3174&39.8780\\
5130380& 19.9830 & 0.0002 & 55009.918408&0.080312&299.1550&40.2881\\
5217781& 564.41 & 0.01 & 55022.074166&0.109098&298.9473&40.3755\\
5310387& 0.4416691 & 0.0000003 & 54953.664664&0.026196&299.8789&40.4824\\
5653126& 38.4969 & 0.0002 & 54985.816152&0.067928&299.7020&40.8963\\
5771589& 10.73914 & 0.00003 & 54962.116764&0.038736&284.5855&41.0096\\
6148271& 1.7853 & 0.0001 & 54966.114761&0.040979&298.2297&41.4598\\
6197038& 9.75171 & 0.00003 & 54961.776580&0.098446&289.3008&41.5265\\
6205460& 3.72283 & 0.00001 & 54956.455222&0.098339&291.9803&41.5594\\
6432059& 0.769740 & 0.000001 & 54964.754300&0.012775&288.0399&41.8184\\
6629588& 2.264471 & 0.000003 & 54966.783103&0.042077&297.7556&42.0091\\
7289157& 5.26581 & 0.00001 & 54969.976049&0.044130&293.9661&42.8373\\
7375612& 0.1600729 & 0.0000001 & 54953.638704&0.008521&295.4670&42.9279\\
7668648& 27.8186 & 0.0001 & 54963.315401&0.054044&286.2770&43.3391\\
7670617& 24.7038 & 0.0001 & 54969.128128&0.044240&287.1845&43.3671\\
7955301& 15.3244 & 0.0001 & 54968.272901&0.420237&290.1863&43.7239\\
8023317& 16.5790 & 0.0001 & 54979.733478&0.052909&289.9703&43.8205\\
8122124& 0.2492776 & 0.0000001 & 54964.612833&0.014220&299.0195&43.9576\\
8365739& 2.38929 & 0.00000 & 54976.979013&0.028860&291.8875&44.3456\\
8758716& 0.1072049 & 0.0000000 & 54953.672989&0.006195&293.8519&44.9494\\
8938628& 6.86222 & 0.00002 & 54966.603088&0.018955&285.6626&45.2177\\
9214715& 265.300 & 0.003 & 55149.923156&0.055997&290.3912&45.6816\\
9715925& 6.30820 & 0.00003 & 54998.931972&0.016542&294.1644&46.4240\\
9834257& 15.6514 & 0.0001 & 55003.014878&0.036964&293.2332&46.6002\\
9944907& 0.613440 & 0.000000 & 54964.826351&0.021577&288.8836&46.8698\\
10014830& 3.03053 & 0.00001 & 54967.124438&0.091901&293.1763&46.9224\\
10223616& 29.1246 & 0.0001 & 54975.144652&0.093130&297.3907&47.2557\\
10268809& 24.7090 & 0.0001 & 54971.999951&0.034276&289.6806&47.3178\\
10319590& 21.3205 & 0.0001 & 54965.716743&0.085213&281.7148&47.4144\\
10743597& 81.195 & 0.001 & 54936.094491&0.029846&296.3823&48.0453\\
10855535& 0.1127824 & 0.0000000 & 54964.629315&0.006374&289.2119&48.2031\\
10919564& 0.4621374 & 0.0000003 & 54861.987523&0.037827&291.5974&48.3933\\
11465813& 670.70 & 0.01 & 55285.816598&0.302424&296.6986&49.3165\\
11558882& 73.921 & 0.001 & 54987.661506&0.044135&291.7399&49.5708\\
11869052& 20.5455 & 0.0001 & 54970.909481&0.022169&294.4019&50.1721\\
12062660& 2.92930 & 0.00001 & 54998.936258&0.043481&292.0709&50.5468\\
 
\hline 
\end{tabular}
\caption{The systems exhibiting Eclipse Depth Variations in the {\it Kepler} sample. Those reported without $\mathrm{BJD}_0$ errors are also heartbeat stars.}
\label{tab:DV} 
\normalsize 
\end{center} 
\end{table}

\subsection{Single Eclipse Events}
Systems exhibiting a primary and/or secondary eclipse but lack a repeat of either one are shown in Table \ref{tab:L}. For these 32 systems no ephemeris, ETV, or period error is determined. These are flagged with the ``L'' (long) flag and are available from the database under \texttt{http://keplerEBs.villanova.edu/search} but are not included in the EB Catalog.

%Show plots of 3625986 or 9466335 when new plots update
\begin{table}[ht!]
\center{}
\begin{center} 
\scriptsize \begin{tabular}{c c c c c c} 
\hline 
\hline 
KIC
&\multicolumn{1}{p{2cm}}{\centering Event Depth\\ (\%)}
&\multicolumn{1}{p{2cm}}{\centering Event Width \\ (days)}
&\multicolumn{1}{p{2cm}}{\centering BJD$_0$ \\ (-2400000)}
&\multicolumn{1}{p{2cm}}{\centering RA($^\circ$)\\ (J2000)}
&\multicolumn{1}{p{2cm}}{\centering Dec($^\circ$) \\ (J2000)}\\ 
\hline
2162635&	0.996&	1.2	&55008	&291.9776	&37.5326\\
3346436&	0.84&	0.9	&55828	&292.4968	&38.4627\\
3625986&	0.75&	12	&55234	&284.7664	&38.7935\\
4042088&	0.983&	0.4	&55449	&286.8895	&39.1074\\
4073089&	0.67&	0.7	&56222	&295.4901	&39.1059\\
4585946&	0.89&	0.8	&54969	&297.3779	&39.6685\\
4755159&	0.9&	0.8	&55104	&294.1376	&39.8821\\
5109854&	0.87&	0.5	&55125	&294.6938	&40.292\\
5125633&	0.92&	0.4	&55503	&298.1539	&40.2474\\
5456365&	0.95&	0.7	&56031	&294.1284	&40.6823\\
5480825&	0.992&	1.5	&55194	&299.5369	&40.6708\\
6751029&	0.88&   0.3	&55139	&281.3179	&42.2645\\
6889430&	0.87&	2	&55194	&298.3608	&42.3808\\
7200282&	0.945&	0.4	&55632	&291.7615	&42.7566\\
7222362&	0.6&	0.6	&55284	&297.5835	&42.7721\\
7282080&	0.965&	0.2	&55539	&291.8008	&42.8907\\
7288354&	0.965&	0.2	&55235	&293.7429	&42.862\\
7533340&	0.98&	0.8	&55819	&293.4575	&43.1115\\
7732233&	0.8&	0.2	&55306	&282.2236	&43.4657\\
7875441&	0.93&	1	&55534	&284.9639	&43.6527\\
7944566&	0.8&	0.5	&55549	&285.1083	&43.7168\\
7971363&	0.965&	1.5	&55507	&295.6284	&43.7571\\
8056313&	0.88&	1.5	&56053	&299.735	&43.8779\\
8648356&	0.98&	0.75&	55357	&299.1302	&44.7447\\
9466335&	0.93&	35	&55334	&292.4365	&46.0956\\
9702891&	0.82&	1.1	&55062	&288.4751	&46.4646\\
9730194&	0.9&	1	&55277	&298.7442	&46.4508\\
9970525&	0.9988&	0.3&	54972	&298.7442	&46.8301\\
10058021&	0.985&	0.3&	55433	&283.7151	&47.0321\\
10403228&	0.955&	2&	    55777	&291.2267	&47.55\\
10613792&	0.975&	0.4&	55773	&296.8537	&47.8918\\
11038446&	0.94&	0.3&	55322	&295.7166	&48.5419\\

\hline 
\end{tabular}
\caption{The systems with no repeating events (Long) in the {\it Kepler} sample. These systems do not have periods}
\label{tab:L}
\normalsize 
\end{center} 
\end{table}

\subsection{Additional Interesting Classes}
The Catalog contains many interesting classes of objects which we encourage the community to explore. Some additional classes, which will only be listed (for brevity), include: total eclipsers, occultation binaries, reflection-effect binaries, red giants, pulsators, and short-period detached systems. These systems can be found using the search criteria provided at \texttt{http://keplerEBs.villanova.edu/search}.

\section{Spectroscopic Follow-up}
\label{sec:spectra}

Modeling light curves of detached eclipsing binary stars can only yield relative sizes in binary systems. To obtain the absolute scale, we need radial velocity (RV) data. In case of double-lined spectroscopic binaries (SB2), we can obtain the mass ratio $q$, the projected semi-major axis $a \sin i$ and the center-of-mass (systemic) velocity $v_\gamma$. In addition, SB2 RV data also constrain orbital elements, most notably eccentricity $e$ and argument of periastron $\omega$. Using SB2 light curve and RV data in conjunction, we can derive masses and radii of individual components. Using luminosities from light curve data, we can derive distances to these systems. When the luminosity of one star dominates the spectrum, only one RV curve can be extracted from spectra. These are single-lined spectroscopic binaries (SB1), and for those we cannot (generally) obtain a full solution. The light curves of semi-detached and overcontact systems can constrain the absolute scale by way of ellipsoidal variations: continuous flux variations due to the changing projected cross-section of the stars with respect to line of sight as a function of orbital phase. These depend on stellar deformation, which in turn depends on the absolute scale of the system. While possible, the photometric determination of the $q$ and $a \sin i$ is significantly less precise than the SB2 modeling, and the parameter space is plagued by parameter correlations and solution degeneracy. Thus, obtaining spectroscopic data for as many EBs as possible remains a uniquely important task to obtain the absolute scale and the distances to these systems.

We successfully proposed for 30 nights at the Kitt Peak National Observatory's 4-m telescope to acquire high resolution ($R \geq 20,000$), moderate signal-to-noise ($S/N \geq 15$) spectra using the echelle spectrograph. To maximize the scientific yield given the follow-up time requirement, we prioritized all targets into the following groups:
\begin{itemize}
\item {\bf Well-detached EBs in near-circular orbits.} These are the prime sources for calibrating the $M$-$L$-$R$-$T$ relationships across the main sequence. Because of the separation, the components are only marginally distorted and are assumed to have evolved independently from one another. Circular orbits indicate that the ages of these systems are of the order of circularization time \citep{zahn:1989, meibom:2005}. Modeling these systems with state-of-the-art modeling codes (i.e.~WD: \citealt{wilson:1979}, and its derivatives; ELC: \citealt{orosz:2000}; PHOEBE: \citealt{prsa:2005}) that account for a range of physical circumstances allows us to determine their parameters to a very high accuracy. These results can then be used to map the physical properties of these main sequence components to the spectral type determined from spectroscopy. \rev{While deconvolution artifacts might raise some concern (cf.~Fig.~\ref{}; left), in practice they become vanishingly small for orbital periods $\gtrsim 1$-d.}

\item {\bf Low-mass main sequence EBs.} \rev{By combining photometric observations during the eclipses and high-$R$ spectroscopy}, we can test the long standing discrepancy between the theoretical and observational mass-radius relations at the bottom of the main-sequence, namely that the observed radii of low-mass stars are up to 15\% larger than predicted by structure models. It has been suggested that this discrepancy may be related to strong stellar magnetic fields, which are not properly accounted for in current theoretical models. All previously well-characterized low-mass main-sequence EBs have periods of a few days or less, and their components are therefore expected to be rotating rapidly as a result of tidal synchronization, thus generating strong magnetic fields. Stars in the binaries with longer orbital periods, which are expected to have weaker magnetic fields, may better match the assumptions of theoretical stellar models. \rev{Spectroscopy can provide evidence of stellar chromospheric activity, which is statistically related to ages, thus discriminating between young systems settling onto the main sequence and the older ones already on the main sequence.}

\item {\bf EBs featuring total eclipses.} A select few EBs with total eclipses allow us to determine the inclination and the radii to an even higher accuracy, typically a fraction of a percent. Coupled with RVs, we can obtain the absolute scale of the system and parameters of those systems with unprecedented accuracy.

\item {\bf EBs with intrinsic variations.} Binarity is indiscriminate to spectral and luminosity types. Thus, components can be main-sequence stars, evolved, compact, or intrinsically variable -- such as pulsators ($\delta$-Sct, RR Lyr, \dots), spots, etc. These components are of prime astrophysical interest to asteroseismology, since we can compare fundamental parameters derived from binarity to those derived from asteroseismic scaling relations \citep{huber:2014}.

\item {\bf EBs exhibiting Eclipse Timing Variations (ETVs).} These variations can be either dynamical or due to the light time effect. The periodic changes of the orbital period are indicative of tertiary components \citep{conroy:2014}. Understanding the frequency of tertiaries in binary systems is crucial because many theories link the third component with the tightening of binary orbits via Kozai-Lidov mechanism and/or periastron interactions. Careful studies of statistical properties of ETVs detected in EBs may shed light on the origins of binarity. In addition, the uncertainty of fundamental parameters derived from multiple stellar systems is an order of magnitude smaller than that of binary stars \citep{carter:2011, doyle:2011}.
\end{itemize}

We acquired and reduced multi-epoch spectra for 611 systems within this program; \rev{typically 2-3 spectra per target were acquired near the quadratures.} The spectra are available for download from the EB Catalog website. \rev{Extrapolating from these systems, we anticipate around 30\% of the {\sl Kepler} sample of EBs presented in the Catalog to be SB2 systems.} The detailed analysis of the acquired spectra is the topic of an upcoming paper (Johnston et al., in preparation).

\section{Catalog Analysis}

\begin{figure}[H]
\includegraphics[width=\textwidth]{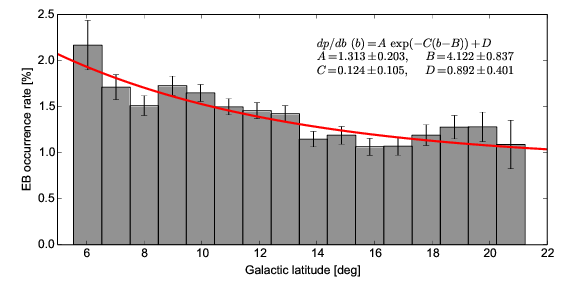} \\
\caption{
\label{fig:latdist}
EB occurrence rate ($dp/db$) as a function of galactic latitude ($b$). Each latitude bin is area-corrected to give true occurrence rates. Uncertainties are estimated as $\sqrt{N_j}/N_{\mathrm{tot},j}$, where $N_j$ is the number of detected EBs and $N_{\mathrm{tot},j}$ is the total number of targets observed by {\sl Kepler} in the $j$-th bin. The solid line represents the exponential fit to the data, with parameters annotated in the Figure. The parameters $A$ and $D$ are occurrence rates, $C$ is the thin-to-thick disk transition value, and $B$ is degrees of galactic latitude. The results of this toy model indicate that the occurrence rate span of EBs ranges between 0.9\% and 2.2\%.
}
\end{figure}

In Papers I and II we noted a non-uniform distribution of eclipsing binary occurrence rates as a function of galactic latitude. This was surprising at first, since the implication that eclipsing binary stars are not uniformly distributed in space is not immediately obvious. Galaxy population observations and models, however, have long predicted this behavior: the stellar population at lower galactic latitudes (thin disk) contains notably younger stars that are on average larger (i.e.~contain more giants in the magnitude-limited sample) than the older, sparser population of the thick disk and halo \citep{prsa:2015}. In consequence, the geometric probability of eclipses increases towards the galactic disk. Fig.\ ~\ref {fig:latdist} depicts this distribution for the complete sample of EBs in the current Catalog. Galactic latitudes are divided into $\sim$1-deg bins and EBs within those bins are counted. To get the area-corrected occurrence rates, their number is divided by the number of all stars observed by {\sl Kepler} within the same bins. The latitude-dependent occurrence rate is immediately obvious. To the $j$-th bin we assigned a corresponding error of $\sqrt{N_j}/N_{\mathrm{tot},j}$, where $N_j$ is the number of detected EBs and $N_{\mathrm{tot},j}$ is the number of all stars in that bin. Unfortunately, {\sl Kepler}'s latitude span is not sufficient to fit an actual galactic population model to the number of EBs detected, so we resort to a toy proxy using a simple exponential fit: $dp/db(b) = A \exp(-C(b-B)) + D$. We performed a least squares fit to derive parameters of the exponential, depicted by a solid line in Fig.~\ref{fig:latdist}. This toy model predicts that the largest expected occurrence rate of EBs is $A + D \sim 2.2\%$, while the smallest expected occurrence rate is $D \sim 0.9\%$. Parameter $B$ drives the galactic latitude offset ($b \sim 4^\circ$) where the exponential rise of EBs in this magnitude-limited sample is still a reasonable approximation, beyond which the disk opaqueness causes the number of EBs to level off and the exponential model to fail. Finally, parameter $C$ provides an estimate of thin-to-thick disk transition, determined predominantly by the decrease in the number of giants in the magnitude-limited sample.

\subsection{Catalog Completeness}

With the primary {\sl Kepler} mission having ended after 4 years of service, the longest orbital periods of EBs in the Catalog are $\sim$1000 days. Catalog completeness at those periods is challenged by our ability to detect every single eclipse event, which can be made difficult by the small eclipse amplitudes, data gaps and other intrinsic and extrinsic contributions to background and noise. Adding to this is the increasingly low probability of eclipses at the larger orbital separations that accompany longer periods, and of course the increasing probability of entirely missed eclipses for orbital periods longer than the observing window. On the other hand, completeness should be $\sim$100\% for short period EBs ($P\sim1$ day) because of the high geometrical probability of eclipses and because even non-eclipsing systems manifest as ellipsoidal variables. The overall Catalog completeness is thus predominantly a function of orbital period and signal-to-noise ratio (SNR) of the eclipses. Here we only estimate completeness and defer an in-depth study that derives the underlying orbital period distribution of all binaries from the Besan\c con model of the Galaxy to Pr\v sa, Matijevi\v c and Conroy (2016; in preparation).

\begin{figure}[H]
\includegraphics[width=\textwidth]{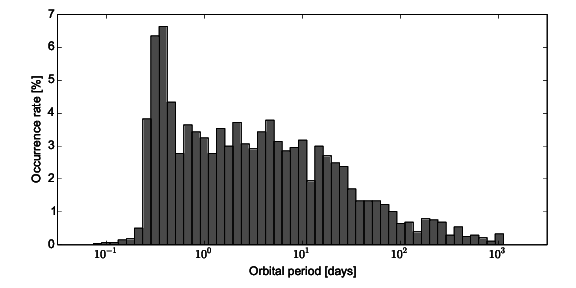} \\
\caption{
\label{fig:perdist}
Distribution of orbital periods of all Cataloged EBs.
}
\end{figure}

To estimate Catalog completeness, we start with the observed period distribution. Fig.~\ref{fig:perdist} depicts the distribution of orbital periods. Two features are particularly interesting: the excess of short period binaries ($P \sim 0.3$ day) and the gradual drop-off of longer period binaries.

The short-period excess is a well-known feature of eclipsing binary stars: at short periods, proximity effects become pronounced, most notably ellipsoidal variations, which enable us to detect binary stars even in the absence of eclipses. This also drives the overall probability of detection sharply upwards at shorter periods. A much broader, less pronounced, longer period peak at $\sim$2-3 days is typically attributed to Kozai-Lidov Cycles and Tidal Friction (KCTF; \citealt{fabrycky:2007}). The Kozai-Lidov mechanism describes interactions with a more distant third companion on an eccentric orbit, where angular momentum is exchanged between the inner binary eccentricity and the outer orbit inclination. When the components of the inner binary are sufficiently close, tidal friction dissipates energy and tightens the pair. The high occurrence rate of stellar triples \citep{gies:2012, rappaport:2013, conroy:2014} gives further credibility to this model.

The gradual drop-off at the long period end is due to two main contributions. The dominant contribution is the geometrical probability of eclipses. Since the eccentricity distribution of {\sl Kepler} EBs can be estimated and the pool of stellar masses and radii can be inferred from the bulk properties of all observed {\sl Kepler} targets, computing this correction is a tractable problem. The second contribution is due to {\sl Kepler}'s duty cycle. The satellite observed a single patch of the sky, but observations were interrupted by regular quarterly rolls and data downlink, and by unexpected events that put the telescope into safe mode. As a result, the actual duty cycle of observations was $\sim$92\%. The dead module and targets observed only for a subset of quarters further impact completeness. We discuss each of these effects in turn and derive corresponding corrections that we use to estimate Catalog completeness. The corrections are depicted in Fig.~\ref{fig:corrections}.

\subsubsection{Geometrical correction}

\begin{figure}[H]
\includegraphics[width=\textwidth]{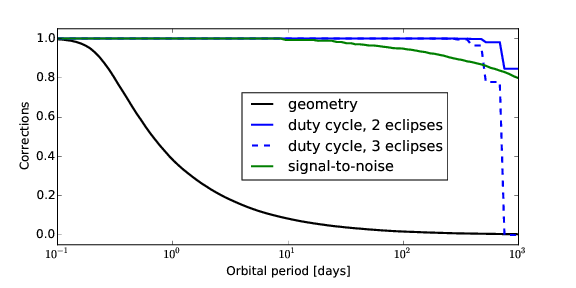}
\caption{
\label{fig:corrections}
Corrections to EB detection rates as a function of orbital period. Geometrical correction is dominant across all orbital periods, while the duty cycle correction becomes important at the long period end. The solid curve for the duty cycle correction is computed by requiring at least two eclipse events to be observed, while the dashed curve shows the duty cycle correction for at least three observed eclipse events. The combined correction is the product of individual corrections.
}
\end{figure}

The probability of eclipses is determined by a simple relation: $\cos i \leq (\rho_1 + \rho_2)/\delta$, where $i$ is orbital inclination, $\rho_1$ and $\rho_2$ are fractional radii, and $\delta$ is the relative instantaneous separation. The right-hand side becomes progressively smaller with increasing orbital periods because of Kepler's 3rd law, whereas the $\cos i$ term is distributed uniformly, so the probability of eclipses drops. For eccentric orbits, the probability of eclipse at superior and inferior conjunction is $p_{\sup, \inf} = (\rho_1 + \rho_2) (1 \pm e \sin \omega) /(1-e^2)$. To compute it, we take all $\sim$200,000 stars observed by {\sl Kepler} and create a pool based on their effective temperatures and surface gravities as reported in the Kepler Input Catalog (KIC; \citealt{brown:2011}). From this pool we draw pairs of stars and use the mass-radius-temperature-log($g$) relationship from \citet{torres:2010} to determine the masses and radii of the drawn stars. We then place these stars in orbits with a predefined orbital period in the $-1 \leq \log P \leq 3$ range, and compute the semi-major axes. We also account for ellipsoidal variability on the short period end and non-zero eccentricity on the long period end. The geometrical correction is depicted in Fig.~\ref{fig:corrections} in black.

\subsubsection{Duty cycle correction}

While Kepler observed approximately 200,000 stars for most of the mission, which targets were observed in which quarters varies at the 10\% level for various reasons, such as compensating for missing modules and variations in target lists over time. To account for these variations and the specific duty cycle, we performed a detailed analysis that leads to a global window function. As an approximation to the search of our algorithm, we use the Transiting Planet Search (TPS) deweighting vector \citep{burke2015} to determine which cadences are actually observed by {\sl Kepler}. This correctly accounts for the exact gaps in the data.

For every target, we use the online stellar table to identify the quarters for which observations were taken. We then scan over periods ranging from 0.5 days to 1500 days, with a trial period corresponding to integer number of {\sl Kepler} cadences (30 minutes). For each period, we also scan over every possible phase (again, in integer cadences) and determine the fraction of phases where Kepler acquired data on this target. This creates an individual window function for each target that corresponds to the probability that an EB with a random phase would have had at least 2 observations as a function of period, which is the minimum requirement to determine the ephemeris of the system. We then sum these individual window functions to determine, for every period, how many Kepler targets would have 2 observations. The correction as a function of period is depicted in Fig.~\ref{fig:corrections} in solid blue.

It is instructive to compare this window function to the theoretical window function if we assume a uniform observing completeness of 92\%. Assuming that an EB was observed for all quarters, the probability of observing any single eclipse is 92\%, and the probability for an eclipse to fall into a data gap is 8\%. The effect is smaller for intermediate period binaries and diminishes for shorter period systems. A binary with an orbital period of $\sim$1000 days will have 2 primary eclipse events in the data, a binary with an orbital period of $\sim$250 days will have 5 eclipse events, and a $\sim$100-day binary will have 14 events. Thus, the probability of a significant fraction of 14 events falling into data gaps is negligibly small, while the effect on 5 events, and especially 2 events, can be dire.

The probability that two or more eclipses were observed is calculated by a binomial expression that depends on the duty cycle and on the number of eclipses, which in turn depends on the orbital period. To compute that probability, it is simpler to compute the probability that no eclipses are detected or that only one eclipse is detected, and take the complement: $p_\mathrm{detection} = 1 - p_0 - p_1 = 1 - (0.08)^N - N (0.92)^1 (0.08)^{N-1}$, where $N$ is the number of eclipses. {\sl Kepler} accumulated 1460 days of data, so $N$ can be written as $1+\mathrm{int}\,(1460/P)$, where $P$ is the orbital period. We expect a cascading correction with a discrete jump at every period that changes the integer division value. This correction is depicted in Fig.~\ref{fig:corrections} in dashed blue.

If we were to require a detection of three eclipses, we would need to subtract another binomial term from the probability equation, $p_2 = 0.5 N(N-1) (0.92)^{N-2} (0.08)^2$, which further reduces the detection rate. This correction is depicted in Fig.~\ref{fig:corrections} in dashed-dotted blue.

\subsubsection{Combined correction}

The combined correction, $\epsilon_\mathrm{C}$, is a product of individual corrections: $\epsilon_\mathrm{C} = \epsilon_\mathrm{G} \epsilon_\mathrm{DC}$. If we knew what the underlying distribution of all (not just eclipsing) binary stars is, then multiplying that distribution with $\epsilon_\mathrm{C}$ would provide us with a theoretical prediction for the observed number of EBs by {\sl Kepler}. The \emph{actual} number of EBs observed by {\sl Kepler}, divided by the \emph{predicted} number of EBs observed by {\sl Kepler}, is then the measure of Catalog completeness.

\begin{figure}[H]
\includegraphics[width=\textwidth]{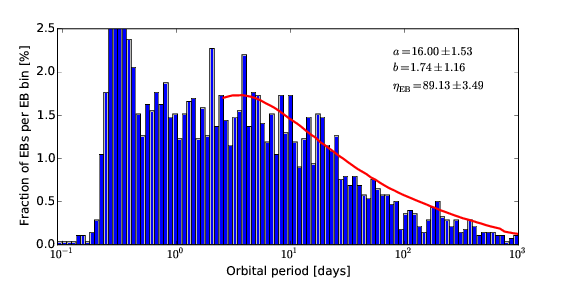}
\includegraphics[width=\textwidth]{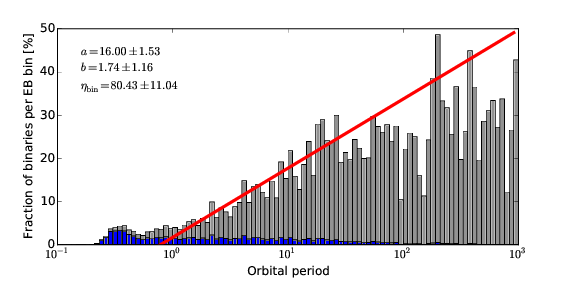}
\caption{
\label{fig:occurrence}
Completeness estimates for the EB Catalog (top), and projected completeness estimates for the binary star population (bottom). The observed distribution of orbital periods is depicted in blue, and the theoretical distribution of orbital periods, derived from a linear model of the underlying binary log-period distribution, is depicted in red. The distribution depicted in gray is the predicted occurrence rate of \emph{all} binary stars based on {\sl Kepler} data, compared to the best-fit linear model.
}
\end{figure}

We do not know the underlying distribution of binary stars, but we can approximate it for the purpose of this estimate. Several works, most notably \citet{duquennoy:1991} and \citet{kroupa:2001}, report a log-gaussian distribution with the peak well in excess of 10,000 days. We are thus sampling a far tail of the left wing, and we can assume that the distribution in the $1.3 \leq \log P \leq 3$ range is locally\footnote{Tidal interaction and the Kozai-Lidov mechanism prohibit this toy model from working below $\log P \sim 1.3$, which is why we do not attempt to model the short period end; since it is reasonable to assume that the Catalog is essentially complete on that end, this deviation from the model bears no significant impact on our completeness result.} linear, $d\mathrm B/d(\log P) = a \log P + b$. By making that assumption, we can then: (1) set $a$ and $b$ to some reasonable initial values; (2) derive the eclipsing binary distribution function $d\mathrm{EB}/d(\log P) = \epsilon_\mathrm C \times d\mathrm B/d(\log P)$; (3) compare this theoretical prediction with the observed distribution, and (4) iterate $a$ and $b$ to obtain the best-fit values on the linear $\log P$ range. Fig.~\ref{fig:occurrence} depicts this comparison: the top plot shows the comparison between the observed EB period distribution (blue bars) and theoretical prediction in the $1.3 \leq \log P \leq 3$ range (solid red line), and the bottom plot shows the comparison between the backwards-projected $d\mathrm{B}/d(\log P) = d\mathrm{EB}/d(\log P) / \epsilon_\mathrm C$ (gray bars) and the assumed underlying linear distribution (red line). To get the completeness estimates, we divide the integral of the observed trend with the integral of the predicted trend. We do not take the short period excess into account because of its extrinsic causes. The annotations in the plots give the best-fit parameters and the derived completeness rates. The completeness of cataloged EBs is $\eta_\mathrm{EB} = 89.1\% \pm 3.5\%$, and completeness of all binaries is $\eta_\mathrm{bin} = 80\% \pm 11\%$.

\section{Summary}

This revision of the Catalog contains new additions consisting of rejected KOIs, previously misidentified false positives, proprietary systems, updated long period systems, EBs identified from other systems, heartbeat stars, and Planet Hunters' systems. Detrended and phased long-cadence data are hosted and available for the public, as well as an updated period for each system adjusted to account for ETVs. We also provide a deconvolved polyfit which serves as a better approximation of the actual light curve and a classification parameter representing the morphology of the phased light curve. A Catalog analysis including a completeness study is provided.

%Double check the things we actually provide; this paragraph and the next one may be able to be merged depending on the answer

An online version of the Catalog is maintained at \texttt{http://keplerEBs.villanova.edu}. This Catalog lists the KIC, ephemeris, morphology, principle parameters, polyfit data, ETV data, raw data, and an array of figures displaying the raw time domain, detrended data, and phased light curves of each system along with period frequency, ETV, and diagnostic analysis plots. The online Catalog also provides a visualization tool to further exploit this data-set. It is our hope that the Catalog will serve the eclipsing binary community as a bridge between the raw \emph{Kepler} data and in-depth scientific modeling.

\section{Acknowledgements}
All of the data presented in this paper were obtained from the Multimission Archive at the Space Telescope Science Institute (MAST). STScI is operated by the Association of Universities for Research in Astronomy, Inc., under NASA contract NAS5-26555. Support for MAST for non-Hubble Space Telescope data is provided by the NASA Office of Space Science via grant NNX09AF08G and by other grants and contracts. Funding for this Discovery Mission is provided by NASA’s Science Mission Directorate. Facility: \emph{Kepler}. Spectroscopic follow-up data are made available through NOAO survey program 11A-0022. This work is funded in part by the NASA/SETI subcontract 08-SC-1041 and NSF RUI AST-05-07542. BQ was supported by an appointment to the NASA Postdoctoral Program at the Ames Research Center, administered by Oak Ridge Associated Universities through a contract with NASA. TSB acknowledges support from ADAP14-0245 and ADAP12-0172. AD has been supported by the Postdoctoral Fellowship Programme of the Hungarian Academy of Sciences, the J\'anos Bolyai Research Scholarship of the Hungarian Academy of Sciences, Lend\"ulet-2009 Young Researchers Programme of the Hungarian Academy of Sciences, the European Community’s Seventh Framework Programme (FP7/2007-2013) under grant agreement no. 269194 (IRSES/ASK) and no. 312844 (SPACEINN).

\bibliography{refs}

\section*{Appendix: The Online Catalog}
The online Catalog provides a searchable database of all the eclipsing binaries found in the entire Kepler data-set along with downloadable content and visualization tools. The online version of the Catalog is currently maintained at \texttt{http://keplerEBs.villanova.edu}. The homepage presents the user with a view of the entire Eclipsing Binary catalog. Along the top tool-bar is the tab, "Search" which provides a page where the Catalog can be filtered, sorted, and exported with a variety of options: ID numbers, eclipse properties, morphology parameter, location elements, ETV properties, effective temperatures, and flags with the ability to export the results to something other than the default HTML table, if desirable. For a complete list of search options, units and explanations, and exporting formats please see the online Help section. 

	For individual EB entries, the online Catalog provides a summary of the EB's physical properties, analytic plots, and provides the time-series data for download in various formats. In addition to the raw data, the polyfit data used to detrend the time-series data and the eclipse timing variation data, along with the suite of diagnostic figures are available for download. The figures provided for each EB entry are: raw data, detrended data, polyfit, ETV, and TPF plots. If available, the individual entry page shows additional spectral observations available for download and any publications concerning that EB entry. 
	
	The Catalog currently maintained at \texttt{http://keplerEBs.villanova.edu} will continue to evolve as additional data is ingested and new techniques enhance our research capabilities. A static version of the online Catalog associated with this paper is maintained at MAST \texttt{https://archive.stsci.edu/}. 

\end{document}